\documentclass[12pt]{article}
\usepackage{epsfig,amssymb}
\usepackage{latexsym}

\newcommand{\bq}{\begin{equation}}
\newcommand{\eq}{\end{equation}}
\newcommand{\bqa}{\begin{eqnarray}}
\newcommand{\eqa}{\end{eqnarray}}
\newcommand{\ben}{\begin{enumerate}}
\newcommand{\een}{\end{enumerate}}
\newcommand{\bc}{\begin{center}}
\newcommand{\ec}{\end{center}}
\newcommand{\bqb}{\begin{eqnarray*}}
\newcommand{\eqb}{\end{eqnarray*}}

\def\gsim{\gtrsim}
\def\lsim{\lesssim}

\def\pr#1#2#3{ Phys. Rev. ${\bf{#1}}$, #2 (#3)}
\def\prl#1#2#3{ Phys. Rev. Lett. ${\bf{#1}}$, #2 (#3)}

\def\np#1#2#3{ Nucl. Phys. ${\bf{#1}}$, #2 (#3)}

\def\epj#1#2#3{ Eur. Phys. J. ${\bf{#1}}$, #2 (#3)}
\def\ijmp#1#2#3{ Int. J. Mod. Phys. ${\bf{#1}}$, #2 (#3)}

\def\aop#1#2#3{Annals of Phys. ${\bf{#1}}$, #2 (#3)}

\def\polon#1#2#3{Acta Phys. Polon. ${\bf{#1}}$, #2 (#3) }

\global\nulldelimiterspace = 0pt

\def\sw{s_W}
\def\cw{c_W}
\def\swsq{s^2_W}
\def\cwsq{c^2_W}
\def\mw{m_W}

\def\mwsq{m_W^2}

\def\tchi{\tilde \chi}

\def\s2beta{{\rm s}_{2 \beta}}

\def\tchi{\tilde \chi}

\begin{document}
\pagenumbering{arabic}
\thispagestyle{empty}
\def\thefootnote{\fnsymbol{footnote}}
\setcounter{footnote}{1}

\begin{flushright}
arXiv:1005.5005 [hep-ph]\\
Corrected version.
 \end{flushright}

\vspace{2cm}
\begin{center}
{\large {\bf  Amplitudes for   $gg  \to VV'$ and their high energy SUSY  constraints.}} \\
 \vspace{1.5cm}
{\large G.J. Gounaris$^a$, J. Layssac$^b$,
and F.M. Renard$^b$}\\
\vspace{0.2cm}
$^a$Department of Theoretical Physics, Aristotle
University of Thessaloniki,\\
Gr-54124, Thessaloniki, Greece.\\
\vspace{0.2cm}
$^b$Laboratoire de Physique Th\'{e}orique et Astroparticules,
UMR 5207\\
Universit\'{e} Montpellier II,
 F-34095 Montpellier Cedex 5.\\
\end{center}

\vspace*{1.cm}
\begin{center}
{\bf Abstract}
\end{center}
We study   how the property of asymptotic  helicity conservation (HCns), expected
for any 2-to-2 process in the minimal supersymmetric model (MSSM),
is realized    in the processes $gg  \to \gamma\gamma,\gamma Z,ZZ,W^+W^-$,
at the 1loop electroweak order and  very high energies.
The violation of this property  for the same process in
 the standard model (SM), is also shown.
This    strengthens the claim that HCns is  specific
to the renormalizable SUSY model,  and  not generally valid in  SM.
HCns  strongly reduces the number of non-vanishing 2-to-2
amplitudes at asymptotic energies in MSSM.

\vspace{01cm}

\def\thefootnote{\arabic{footnote}}
\setcounter{footnote}{0}
\clearpage

\section{Introduction}

\hspace{0.7cm} Supersymmetry confers the remarkable Helicity Conservation (HCns)
property to all the $2\to 2$  amplitudes, at high energy and fixed angle.
This was established to all perturbation orders in \cite{heli1, heli2},
for the minimal supersymmetric model (MSSM); provided the energy is so high, that all SUSY
masses can be neglected. Renormalizability is essential in proving HCns,
since any   known anomalous  coupling  would violate it \cite{Kasimierz}.

 More explicitly,  HCns  states that for any process
\bq
a_{\lambda_a}+b_{\lambda_b} \to c_{\lambda_c}+d_{\lambda_d} ~~, \label{gen-process}
\eq
with  $\lambda_j$ denoting  the particle\footnote{Scalar, fermion or gauge boson.} helicity,
all non-vanishing amplitudes at asymptotic energies and fixed angles,
must be helicity conserving (HC), satisfying
\bq
\lambda_a+\lambda_b =\lambda_c+\lambda_d  ~~. \label{HC-constraint}
\eq
Consequently, all  helicity  amplitudes that violate (\ref{HC-constraint}),
must vanish at energies much larger than the SUSY scale, at fixed angles.
Such amplitudes are termed as helicity violating (HV) ones. Evidently,
HCns   drastically  reduces
the number of the relevant asymptotic amplitudes in SUSY.

For processes
involving external gauge bosons, the HCns theorem appears  striking even
at the Born level, where huge cancelations among the various diagrams contrive for its validity.

If the Born contribution to a process is non-vanishing, then HCns is
 approximately correct in the standard model (SM)
 also, up to the 1loop leading-log order \cite{heli1, heli2, Kasimierz, ugdW}.
 By this we mean that in such an SM case,  the HV amplitudes,  although not necessarily vanishing,  are usually much smaller than the HC ones \cite{MSSMrules1,MSSMrules2, MSSMrules3}.

An approximate validity of HCns in SM, has also been observed at 1loop,
in $\gamma \gamma \to \gamma \gamma, ~\gamma Z, ~ZZ$, where the $W$-loop
contribution   is so overwhelming,
that it renders the difference between SM and MSSM  tiny
\cite{gamgamgamgam, gamgamgamZ, gamgamZZ, gamgamZZ2}. Thus the exact HCns validity in MSSM,
forces its approximate validity in SM.

Since the HCns proof in \cite{heli1, heli2} was done by neglecting all SUSY soft breaking masses
and the $\mu$-term, the worry that terms involving ratios of masses
might  possibly invalidate the general proof in \cite{heli1, heli2}, comes   to mind.
The only feasible way to address such worries, is through detail 1loop
calculations, keeping all mass terms. The realization of
helicity conservation  as the energy increases, may then be investigated.

This was first investigated in   detail 1loop
electroweak  (EW) calculations  for
   $ug\to d W^+$ \cite{ugdW} and\footnote{$\tchi_j^+$ describes a chargino.}
   $ug\to \tilde d_L \tchi_j^+$ \cite{ugsdWino}, which confirm  HCns  for these processes,
 and leads to interesting  asymptotic  SUSY relations among the corresponding
 unpolarized  differential cross sections \cite{ugdW, ugsdWino}.
 These cross section relations  turn out to be  approximately correct
 even at LHC-like energies, for a wide range  of SUSY benchmarks
 \cite{SPA1, SPA2, other-bench1,  other-bench2,  other-bench3}.

Subsequently, the study of  the  gluon-fusion processes  $gg\to HH', ~VH$,
  to 1loop EW order,  was done,  where $H,~H'$
 are Higgs or Goldstone bosons  and $V=\gamma,Z,W^\pm$ \cite{ggHHVH}.
The interest   here is  that there  is  neither  a Born  nor a
 $W$-loop contribution, like the one dominating the  $\gamma\gamma$ processes
 mentioned  above.
So,  there was   a chance to  find   processes  where  HCns is strongly violated in SM,
 while of course always obeyed in MSSM \cite{ggHHVH}.
 Such properties were indeed found in
 $gg \to H_{SM} H_{SM}$ involving the SM Higgs particles;
as well as  in the Goldstone involving processes $gg \to G^0G^0, ~G^+G^-$, and
 $gg \to Z_LG^0,~ W^+_LG^-$, related through the equivalence theorem
 \cite{equiv1,equiv2, equiv3, ggHHVH}.\\

The purpose of the present work is to  explore the validity of HCns in MSSM, and
its violation in SM, in the  $gg  \to VV'$ processes for $VV'=\gamma\gamma$, $\gamma Z$,
$ZZ$, $W^+W^-$, calculated at the  1loop EW order.
 In this case, we should   meet  HC amplitudes   involving transverse,
as well as longitudinal, gauge bosons, and hopefully  find
many new  instances, where HCns is realized  in MSSM, and strongly violated in SM.

In achieving this, we construct simple  analytic expressions for all the 1loop
helicity amplitudes at asymptotic energies,
in either SM or MSSM.  At such high energies,  the  amplitudes only  depend
on the gauge couplings,
and they either go to generally  angle-dependent "constants", or vanish.

In MSSM, in agreement with HCns,  only the  HC amplitudes survive asymptotically;
while the  HV amplitudes  go to zero,  due to  spectacular cancelations among
the  quark and squark contributions.

In  SM though, all   HV amplitudes with the  final gauge bosons being
either both transverse or both longitudinal,
go asymptotically to constants, similar in magnitude to the HC amplitudes.
Only those where one of the final gauge bosons is transverse and the other longitudinal,
continue to vanish in SM.

Thus $gg  \to VV'$ provide a rich set of examples
where HCns is strongly violated in SM, while obeyed in MSSM.
This further strengthens the claim that HCns is  a genuine SUSY property \cite{heli1, heli2}.
Such processes have been studied before,
but the helicity properties of the amplitudes were not noticed
\cite{Berger}.\\

 A Fortran code supplying the complete 1llop helicity amplitudes for all
the  $gg\to VV'$ processes,  in either SM or  MSSM, at any c.m. energy and  angle,
   is released in \cite{code}. Using these,  we present  illustrations in  SM and  MSSM.
The  implied  cross sections induced from  the
  $gg \to \gamma\gamma, ~ \gamma Z, ~ZZ, ~ W^+ W^- $ contributions
   at LHC, are also shown, in MSSM  and  SM. If the squarks lie   in the TeV region,
   the MSSM  changes to the $gg \to ZZ, ~ W^+ W^- $ cross sections, relative to the SM
   ones, are found at the (20-30)\% level for TeV c.m. energies.
   We argue that such effects may  be observable.  \\

The contents of the paper are the following.
Sect.2 summarizes the complete  one loop calculation of
 $gg\to VV'$.
In Sect.3 we present the simple analytic expressions for  the  high
energy limit of the various  amplitudes, in both MSSM and SM.
In Sect.4, the Fortran   code is introduced  and various illustrations for the
amplitudes and cross sections are given.  A summary of the  results
and  possible future developments are given in Sect.5.\\

\section{The $gg\to VV'$ diagrams and amplitudes.}

The gluon-fusion processes to two electroweak (EW) vector bosons addressed here are
\bq
g^a(l,\mu) g^b(l',\mu')   \to  V(p,\tau) V'(p',\tau') ~~, \label{processes}
\eq
where $(VV'=\gamma \gamma, ~\gamma Z,~ZZ, W^+W^-)$. Here
 $(l,l',p,p')$ describe the momenta of incoming gluons and outgoing vector bosons,
while $(\mu, \mu', \tau, \tau')$ denote their corresponding helicities.
The indices $a,b$ in (\ref{processes}) denote the gluon  color, so that non-vanishing amplitudes could only appear  for $a=b$.

Following the standard Jacob-Wick conventions \cite{JW}, the helicity amplitudes
 for these processes are denoted as
\bq
F^{VV'}_{\mu\mu'\tau\tau'}(\sqrt{s}, \theta) ~~, \label{Fdef}
\eq
where    $(\sqrt{s}, \theta)$ describe  the subprocess  c.m. energy and angle, while  $\sqrt{S}$  denotes
 the total LHC c.m. energy. A  gluon color factor  $\delta^{ab}$ is always
removed from  the $F$-amplitude, defined so that the  phase of $iF$ coincides  with
 the S-matrix phase.

For the   gluons we always have  $(\mu,\mu'=\pm 1)$.
Depending on the final vector-boson-helicities,
the  amplitudes where both $VV'$  are transverse are referred to as
transverse-transverse (TT), those where one is transverse and the other longitudinal are
called LT or TL amplitudes,  while  those with  $\tau=\tau'=0$ are called   LL amplitudes.

\begin{figure}[p]
\[
\hspace{-3.8cm}
\epsfig{file=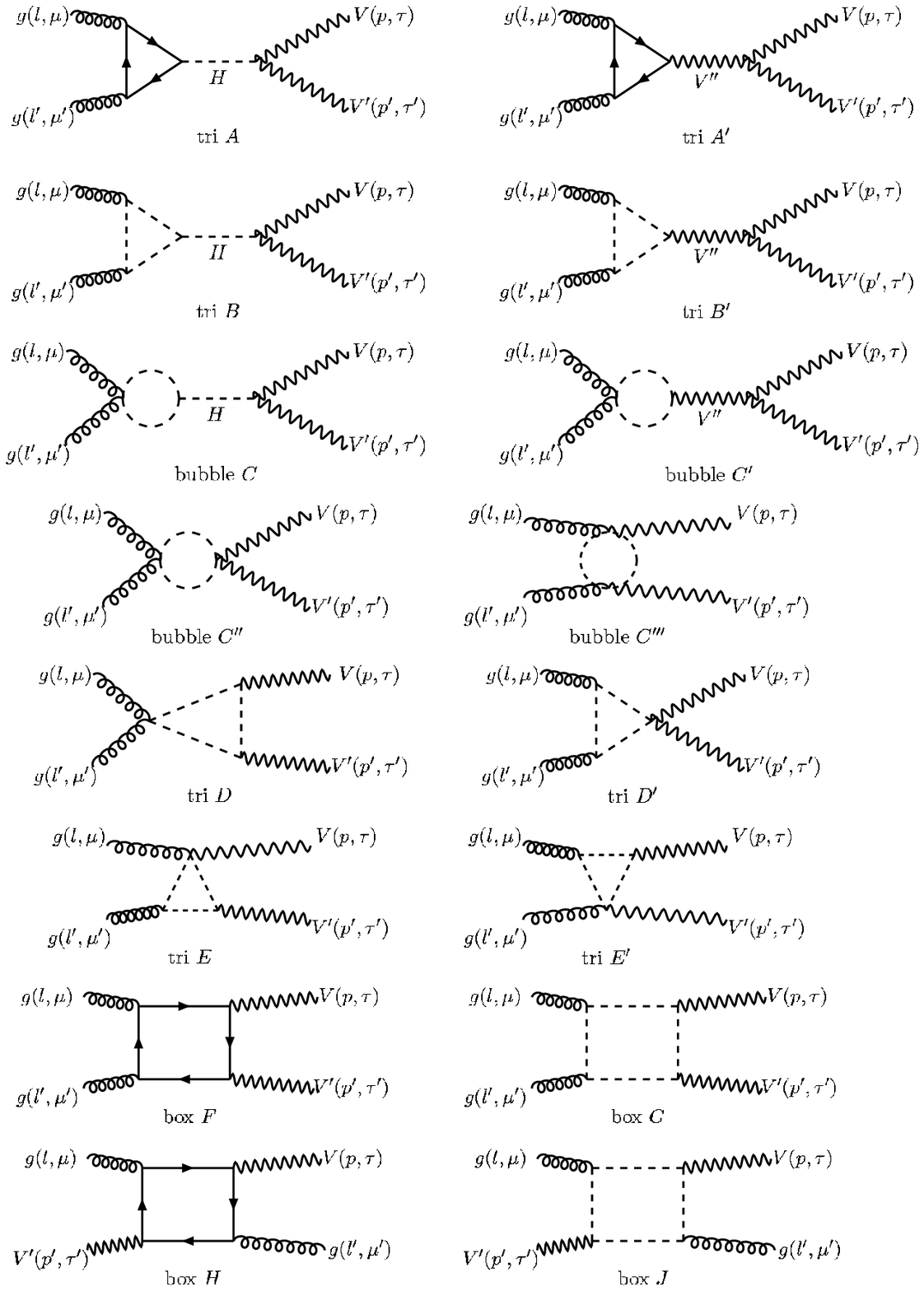,height=21.cm, width=20cm}
\]
\caption[1]{Independent EW contributing graphs to $gg\to VV'$.
Full internal lines describe quark exchanges, while
broken lines denote squark exchanges. The Higgs $H$ exchanges in the graphs tri-$A$,
tri-$B$ and tri-$C$, are also denoted by broken lines. $V,V',V''$ denote the
EW gauge bosons.}
\label{diagrams-fig}
\end{figure}

As already said, there is no Born EW contribution to the processes in
(\ref{processes}). Non-vanishing contributions first arise at 1loop,
where the needed  independent graphs  for MSSM are shown in Fig.\ref{diagrams-fig}.
A subset of these graphs involving only quark exchanges, and where Higgs
exchange corresponds to $H_{SM}$, determines the SM result.

The squark-exchange bubbles $C,~C',~ C''$ and the triangle  $D$ in  Fig.\ref{diagrams-fig}
are by definition symmetric under the exchange of the two initial gluons.
All others must be $gg$-symmetrized
by adding the corresponding gluon symmetrization  (gSYM) contribution
 though  the interchange
\bq
\mu \leftrightarrow \mu' ~~~,~~~ \cos\theta \leftrightarrow -\cos\theta  ~~~,~~~
\sin\theta \leftrightarrow -\sin\theta  ~~~.\label{SYM}
\eq
Taking gSYM into account, we immediately see  that charge conjugation forces
the triangles $A'$ and  $B'$, and the bubble $C'$ to vanish separately.

Since in the graphs of  Fig.\ref{diagrams-fig}, we never exchange $V$ and $V'$,
the antiquark or  antisquark loop
contributions are  sometimes  physically distinct  and must
 be added. This is the case  for   the $F$-box.
For the $H$-box though,
the quark and antiquark loops have the  \underline{same} physical
origin, and therefore only one must be considered.
The calculation
of the $F$ and $H$ boxes, was particularly laborious, since it needed
traces of  eight gamma matrices, with or without
an additional\footnote{Only needed for the $W^+W^-$ amplitudes.}
$\gamma_5$, leading to  boxes with at most two different masses along the internal lines.

Correspondingly, the squark and antisquark loops provide
physically distinct contributions, to
the triangles $D,~E,~E'$ and the box $G$,
and should both be included. On the contrary, for the box $J$, only the squark loop
must be taken into account, since the antisquark loop describes  the same effect.\\

Adding all non-vanishing graphs in  Fig.\ref{diagrams-fig}
and performing the appropriate  gSYM,
all divergences  cancel out. This is realized as follows.
Triangle $A$ and the sum of the graphs  $B+C$ are  convergent.
Quark boxes $F,H$ are divergent, but $F+H$ is convergent.
Squark graphs $C'',D'$ are divergent, but $C''+D'$ is convergent.
Squark triangles  $D, E, E'$ and the bubble $C'''$ are  divergent,  but the divergence
of their  sum  cancels against the divergence of the  boxes
$G+J$.

The calculation of the graphs in  Fig.\ref{diagrams-fig} was done analytically, expressing
the  helicity amplitudes defined in (\ref{Fdef}), in terms of Passarino-Veltman
functions \cite{Veltman}, multiplied
by forms depending  on the particle-helicities and the c.m. energy and angle..
For simplicity, we have neglected  CP violation in either SM or MSSM.
The amplitudes   satisfy the following symmetries \cite{JW}:

Bose-statistics  for the initial gluons gives
\bq
F^{VV'}_{\mu\mu'\tau\tau'}(\theta) =(-1)^{\tau-\tau'}F^{VV'}_{\mu'\mu\tau\tau'}(\pi-\theta)
~~, \label{Bose-gg}
\eq
while  $VV'$-exchange implies
\bq
F^{VV'}_{\mu\mu'\tau\tau'}(\theta) =F^{V'V}_{\mu\mu'\tau'\tau}(\pi-\theta)
~~, \label{Bose-VVp}
\eq
for all $VV'$-channels.

CP invariance   for $VV' =\gamma \gamma,~ \gamma Z,~ZZ$, gives
\bq
F^{VV'}_{\mu\mu'\tau\tau'}(\theta) =(-1)^{\tau-\tau'}
F^{VV'}_{-\mu, -\mu', -\tau, -\tau'}(\theta) ~~, \label{CP-V-neutral}
\eq
while for the charged channel $W^+W^-$ we get
\bq
 F^{W^+W^-}_{\mu\mu'\tau\tau'}(\theta)=
(-1)^{\tau-\tau'}F^{W^+W^-}_{\mu' \mu \tau \tau'}(\pi-\theta)
= F^{W^+W^-}_{-\mu', -\mu, -\tau', -\tau}(\theta) ~~.  \label{CP-WW}
\eq
In addition, at the 1loop level, we find
\bq
 F^{W^+W^-}_{+++-}(\theta)=F^{W^+W^-}_{++-+}(\theta) ~~. \label{1loop-WW}
\eq

Combining (\ref{Bose-gg}, \ref{Bose-VVp}) for the $\gamma\gamma$-channel,
we get\footnote{In addition to it,  there exists also a photon-gluon symmetry,
because of their common $\gamma_\mu$-coupling.}
\bq
F^{\gamma \gamma}_{\mu\mu'\tau\tau'}(\theta) =F^{\gamma \gamma}_{\mu'\mu\tau'\tau}(\theta)
~~. \label{gamma-gamma-sym}
\eq
Correspondingly, for $gg\to ZZ$,
the combination of (\ref{Bose-gg}, \ref{Bose-VVp}, \ref{CP-V-neutral}),
implies
\bqa
&& F^{ZZ}_{\mu\mu'\tau\tau'}(\theta) =(-1)^{\tau-\tau'}F^{ZZ}_{\mu'\mu\tau\tau'}(\pi-\theta)
~~, \nonumber \\
&& F^{ZZ}_{\mu\mu'\tau\tau'}(\theta) =F^{ZZ}_{\mu\mu'\tau'\tau}(\pi-\theta) ~~, \nonumber \\
&& F^{ZZ}_{\mu\mu'\tau\tau'}(\theta) =(-1)^{\tau-\tau'}F^{ZZ}_{\mu'\mu\tau'\tau}(\theta) ~~,
\nonumber \\
&& F^{ZZ}_{\mu\mu'\tau\tau'}(\theta)
=(-1)^{\tau-\tau'}F^{ZZ}_{-\mu -\mu'-\tau -\tau'}(\theta) ~~.
\label{ZZ-sym}
\eqa

These symmetries constrain the  number of independent helicity amplitudes
of  the various processes $gg \to VV'$. Below we enumerate these independent  amplitudes,
 conveniently separating them in  three classes. The helicity conserving (HC) amplitudes
 that respect (\ref{HC-constraint}); the helicity violating TT and LL
 amplitudes,  $(HV_{TT,LL})$  that violate (\ref{HC-constraint});
 and finally the transverse-longitudinal $HV_{TL}$
 and   $HV_{LT}$, which also violate  (\ref{HC-constraint}).

As a result, for $gg\to \gamma \gamma$, there are 4
 independent helicity amplitudes, namely
\bqa
HC \Rightarrow && F^{\gamma \gamma}_{++++}(\theta)~~,~~
F^{\gamma \gamma}_{+-+-}(\theta)=F^{\gamma \gamma}_{+--+}(\pi-\theta)~~,~~
\nonumber \\
HV_{TT} \Rightarrow && F^{\gamma \gamma}_{++--}(\theta)~~,~~
 F^{\gamma \gamma}_{+++-}(\theta)=F^{\gamma \gamma}_{++-+}(\theta)=
F^{\gamma \gamma}_{+---}(\theta)=F^{\gamma \gamma}_{+-++}(\theta)~~.
\label{F-ind-gammagamma}
\eqa

Correspondingly, for $gg\to \gamma Z$, we take as independent the 9 amplitudes
\bqa
 HC \Rightarrow && F^{\gamma Z}_{++++}(\theta)~,~
  F^{\gamma Z}_{+-+-}(\theta)=F^{\gamma Z}_{+--+}(\pi-\theta)~,~ \nonumber \\
HV_{TT} \Rightarrow && F^{\gamma Z}_{+++-}(\theta)~~,~~ F^{\gamma Z}_{++-+}(\theta)~,~
F^{\gamma Z}_{++--}(\theta)~~,
F^{\gamma Z}_{+-++}(\theta)=F^{\gamma Z}_{+---}(\pi-\theta)~, \nonumber \\
HV_{TL} \Rightarrow && F^{\gamma Z}_{+++0}(\theta)~~,~~ F^{\gamma Z}_{++-0}(\theta)~,~
F^{\gamma Z}_{+-+0}(\theta)=F^{\gamma Z}_{+--0}(\pi-\theta)~. \label{F-ind-gammaZ}
\eqa

 For $gg \to ZZ$,   (\ref{ZZ-sym}) implies 10  independent amplitudes taken as
\bqa
HC \Rightarrow && F^{ZZ}_{++++}(\theta)~~,~~
F^{ZZ}_{+-+-}(\theta)=F^{ZZ}_{+--+}(\pi-\theta)~~,~~
 F^{ZZ}_{+-00}(\theta)~~, \nonumber \\
HV_{TT,LL} \Rightarrow && F^{ZZ}_{+++-}(\theta)~~,~~ F^{ZZ}_{+-++}(\theta)~~,~~
F^{ZZ}_{++--}(\theta)~~,~~
F^{ZZ}_{++00}(\theta)~~\nonumber \\
HV_{TL} \Rightarrow && F^{ZZ}_{+++0}(\theta)~~,~~ F^{ZZ}_{+-+0}(\theta)~~,~~
 F^{ZZ}_{++-0}(\theta)~~.
\label{F-ind-ZZ}
\eqa

 Finally for $gg\to W^+W^-$, the combination of
 (\ref{Bose-gg}, \ref{CP-WW}, \ref{1loop-WW}), implies
 14  independent amplitudes for which we  take
\bqa
HC \Rightarrow && F^{W^+W^-}_{++++}(\theta)~~,~~
 F^{W^+W^-}_{+-+-}(\theta)~~,~~ F^{W^+W^-}_{+--+}(\theta)~~, ~~
F^{W^+W^-}_{+-00}(\theta)~~, \nonumber \\
HV_{TT,LL} \Rightarrow && F^{W^+W^-}_{+++-}(\theta)~~,~~
F^{W^+W^-}_{+-++}(\theta)~~,~~ F^{W^+W^-}_{++--}(\theta)~~,~~  F^{W^+W^-}_{++00}(\theta)~~,
\nonumber \\
HV_{TL} \Rightarrow  &&   F^{W^+W^-}_{+++0}(\theta)~~,~~  F^{W^+W^-}_{++-0}(\theta)~~,~~
F^{W^+W^-}_{+-+0}(\theta)~~,~~    F^{W^+W^-}_{+--0}(\theta) ~~, \nonumber \\
HV_{LT} \Rightarrow &&   F^{W^+W^-}_{++0+}(\theta) ~~,~~
     F^{W^+W^-}_{++0-}(\theta)~~.  \label{F-ind-WW}
\eqa \\

The analytic results derived from Fig.\ref{diagrams-fig},
are used to construct the Fortran code  released together with this paper, which calculates
all helicity amplitudes.\\

\section{Asymptotic behavior  of  $F_{\mu \mu'\tau \tau'}$ in SM and MSSM.}

In this Section we give the analytic asymptotic expressions for all 1loop
independent helicity amplitudes
for the processes in (\ref{processes}). These are  valid at very high energies
and fixed angles, in both SM and MSSM.
For MSSM  this means   $\sqrt{s}\gg M_{SUSY}$, with $M_{SUSY}$
describing the scale of the squark  masses, and satisfying $M_{SUSY}\sim 1 ~{\rm TeV}$,
for the benchmarks \cite{SPA1, SPA2,  other-bench1,  other-bench2, other-bench3}.
For SM, $\sqrt{s}\gg \mw$ is  sufficient.

It turns out that in MSSM,  all HV amplitudes ($HV_{TT}$, $HV_{LL}$ and $HV_{TL}$)
indeed vanish asymptotically,  while the   HC ones  tend to   non-vanishing,
angle-dependent "constants".

 In SM,  the   HC amplitudes continue to behave like angle-dependent
 "constants" asymptotically.
 But in this case,  the  $HV_{TT}$  and $HV_{LL}$ amplitudes are also non-vanishing
 asymptotically, behaving like real constants, independent of the angle.
 The   $HV_{TL}$-amplitudes continue to tend to   a vanishing
 limit  in\footnote{By the equivalence theorem,
 this result is identical to the asymptotic vanishing of the $gg\to V G$ amplitudes found
 in \cite{ggHHVH} and induced by a  mass suppression
 appearing  already at the trace computation.} SM (as they were also doing in MSSM).

So we only need to discuss the  "constant" limits of the TT
and LL  amplitudes. These may be directly obtained from the analytical results for the graphs
in Fig.\ref{diagrams-fig} in terms of the Passarino-Veltman functions \cite{Veltman},
 using the asymptotic expressions  given e.g in \cite{techpaper}.
Alternatively, for $VV'=\gamma\gamma, ~\gamma Z, ~ZZ$, these may be  obtained from the
$\gamma\gamma$-fusion results of \cite{gamgamZZ, gamgamgamgam, gamgamgamZ}.
For the $WW$ case though, some extra work had to be done, particularly concerning the
proof that the $\gamma_5$ contribution of the  $F+H$ boxes in Fig.\ref{diagrams-fig},
indeed vanishes asymptotically. Having done this,
the  values of the aforementioned  limits, (which fully agree with
the results of the 1loop code \cite{code}) have   been obtained. They are presented below,
separating the TT and LL amplitudes:   \\

\noindent
The \underline{TT-amplitudes}\\
Their  general structure for all  $gg\to VV'$ processes  in MSSM has the form \cite{gamgamZZ}
\bq
F_{\mu\mu'\tau\tau'} \to  {\alpha\alpha_s\over2} C_{VV'q} [f^q_{\mu\mu'\tau\tau'}
+ 2 f^{\tilde q}_{\mu\mu'\tau\tau'} ] ~~, \label{Fasym1}
\eq
where $f^q$  and  $f^{\tilde q}$ denote respectively  the $V$-independent part
of the asymptotic quark and squark box loop contributions. Expression (\ref{Fasym1})
may also be used for SM calculations, provided the $f^{\tilde q}$-term is dropped.

 The relative magnitudes of the various processes
in (\ref{Fasym1}) is solely determined by the relevant couplings \cite{gamgamZZ}
\bqa
C_{VV'q} &= & \frac{1}{2}\sum_{\rm flavors}
\left [ g^L_{Vq}g^L_{V'q}+ g^R_{Vq}g^R_{V'q} \right ]
~~,  \label{Casym1}
\eqa
where   the summation is  over all quark  flavors.
For three quark-generations, we thus find
\bq
 C_{ZZq}   =  {(9-18s^2_W+20s^4_W)\over 12 \swsq \cwsq }  ~, ~
 C_{\gamma Z q}   =    {(9-20s^2_W)\over 12 \sw\cw }   ~, ~
 C_{\gamma \gamma  q}   =  \frac{15}{9} \nonumber  ~, ~
C_{WW q}   =  \frac{3}{4\swsq}   ~, \label{Casym2}
\eq
for $gg\to ZZ, ~ \gamma Z,~ \gamma \gamma$, and $W^+W^-$, respectively.\\

The explicit expressions for the $f^q,~f^{\tilde q}$ in (\ref{Fasym1}) are
 \cite{gamgamZZ}
\bq
f^q_{\mu\mu'\tau\tau'}=-2A^S_{\mu\mu'\tau\tau'}+\delta_{\mu\mu'\tau\tau'}~~~, ~~~
f^{\tilde q}_{\mu\mu'\tau\tau'}=A^S_{\mu\mu'\tau\tau'}~~, \label{fqftilq}
\eq
where the only non-vanishing contributions are
\bqa
&& A^S_{++++}=A^S_{----}=4-{4ut\over s^2}\left [\ln^2\Big |{t\over u}\Big |
+ \pi^2\right ]+{4(t-u)\over s}\ln \Big |{t\over u} \Big | ~~, \nonumber \\
&& A^S_{+-+-}=A^S_{-+-+}=4-{4st\over u^2}
\left [\ln^2 \Big |{s\over t} \Big | -2i \pi \ln \Big |{s\over t}\Big | \right ]
+{4(s-t)\over u} \left [ln \Big |{s\over t}\Big |-i\pi \right ] ~~, \nonumber \\
&& A^S_{+--+}=A^S_{-++-}=4-{4su\over t^2}
\left [\ln^2 \Big |{s\over u}\Big | -2i \pi \ln \Big |{s\over u}\Big |\right ]
+{4(s-u)\over t} \left [ln \Big|{s\over t}\Big |-i\pi \right ]~~, \label{ASHC} \\
&& \delta_{++++}=\delta_{----}=-4\left[\ln^2\Big |{u\over t}\Big |+\pi^2 \right ] ~~,
\nonumber  \\
&& \delta_{+-+-}=\delta_{-+-+}=\delta^t \equiv -4 \left [\ln^2\Big |{s\over t}\Big |
-i~2\pi \ln \Big |{s\over t}\Big |\right] ~~, \nonumber \\
&& \delta_{+--+}=\delta_{-++-}=\delta^u \equiv -4 \left [\ln^2\Big |{s\over u}\Big |
-i~2\pi \ln \Big |{s\over u}\Big |\right ] ~~, \label{deltaHC}
\eqa
for the $HC_{TT}$ amplitudes, and
\bqa
&&A^S_{+++-}=A^S_{+-++}=A^S_{++--}=A^S_{++-+}=A^S_{+---}=A^S_{---+}=A^S_{-+--}
\nonumber\\
&& =A^S_{--++}=A^S_{--+-}=A^S_{-+++}~=~-4  ~~, \label{ASHV}
\eqa
for  the $HV_{TT}$ amplitudes. Note that the substitution of ~$t=-s (1-\cos\theta)/2$~
and ~$u=-s (1 +\cos\theta)/2$~ in  the asymptotic relations (\ref{ASHC},
\ref{deltaHC}),  provides  consistent  equivalent expressions
in terms of  $\theta$.

Substituting (\ref{Casym2}-\ref{ASHV}) in (\ref{Fasym1}),
 one observes that in \underline{MSSM}, the squark contributions $f^{\tilde{q}}=A^S$
cancels out  the $A^S$ part of the quark boxes, so that non-vanishing contributions
can only arise from  the $\delta$-terms in (\ref{deltaHC}).
Therefore, only
the  HC amplitudes  $F_{+-+-}$, $F_{+--+}$, $F_{-+-+}$, $F_{-++-}$, $F_{++++}$
and $F_{----}$ survive asymptotically, acquiring energy-independent,  but
 angle-dependent  values. A compact form of them may be written as
 \bqa
&& F(gg\to ZZ)^{\rm as}_{\mu\mu' \tau\tau'} =\alpha\alpha_s
{(9-18s^2_W+20s^4_W)\over 24 s^2_Wc^2_W}
\delta_{\mu\mu' \tau\tau'} ~~, \nonumber \\
&& F(gg\to \gamma Z)^{\rm as}_{\mu\mu' \tau\tau'} = \alpha\alpha_s
{(9-20s^2_W)\over 24 s_Wc_W}
\delta_{\mu\mu' \tau\tau'} ~~, \nonumber \\
&& F(gg\to \gamma\gamma)^{\rm as}_{\mu\mu' \tau\tau'} = \alpha\alpha_s \, {5\over 6}
\, \delta_{\mu\mu' \tau\tau'} ~~, \nonumber \\
&& F(gg\to W^+W^-)^{\rm as}_{\mu\mu' \tau\tau'} = \alpha\alpha_s \, {3\over 8 s^2_W}
\, \delta_{\mu\mu' \tau\tau'}~~,  \label{HC-VVTTasym}
\eqa
in terms of (\ref{deltaHC}).

On the contrary, in \underline{SM} there are no squarks, so that  only the quark box
contribution $f^q=-2A^S+\delta$ must be retained in (\ref{Fasym1}, \ref{fqftilq}).
Using then (\ref{ASHC}-\ref{ASHV}), we obtain
non-vanishing asymptotic limits for \underline{all} the $HC_{TT}$ and
the $HV_{TT}$ amplitudes.
The $HC_{TT}$ limits continue to depend on the angle; see (\ref{ASHC}, \ref{deltaHC}).
But    $HV_{TT}$ amplitudes in SM, all tend to the \underline{same} constant value,
independent of the scattering angle $\theta$; see(\ref{ASHV}).   \\

\noindent
The \underline{LL-amplitudes}\\
These asymptotic limits only concern the  $gg\to Z_LZ_L, ~W_L^+ W_L^-$ amplitudes,
which should be equivalent to $gg\to G^0G^0, ~G^+G^-$ respectively \cite{equiv1, equiv2, equiv3}.

In \underline{SM}, non vanishing  "constant" limits are found for both, the  $HC_{LL}$
amplitudes \cite{gamgamZZ, ggHHVH}
\bqa
&& F^{ZZ}_{+-00}(\theta)=F^{ZZ}_{-+00}(\theta)\to  \alpha\alpha_s
{(m^2_t+m^2_b)\over 16 s^2_W m^2_W} \Big \{\delta^t{(1-\cos\theta)\over 1+\cos\theta}+
\delta^u{(1+\cos\theta)\over1-\cos\theta}\Big \} ~~, \nonumber \\
&& F^{W^+W^-}_{+-00}(\theta)
=F^{W^+W^-}_{-+00}(\pi-\theta)\to {\alpha\alpha_s\over 8 \swsq \mwsq }
\Big \{\delta^t  {m_b^2(1-\cos\theta)\over1+\cos\theta}+
\delta^u   {m^2_t(1+\cos\theta) \over 1-\cos\theta} \Big \}, \label{HCLLasym}
\eqa
where $\delta^t, \delta^u$  are defined in (\ref{deltaHC}); and the $HV_{LL}$ amplitudes
\bqa
&& F^{ZZ}_{++00}(\theta)=F^{ZZ}_{--00}(\theta) \to -\alpha\alpha_s
{(m^2_t+m^2_b)\over s^2_W m^2_W}
~~ \nonumber \\
&& F^{WW}_{++00}(\theta)=F^{WW}_{--00}(\pi-\theta)
\to  -\alpha\alpha_s{(m^2_t+m^2_b) \over s^2_W m^2_W} ~~. \label{HVLLasym}
\eqa

In  \underline{MSSM},   the  $HV_{LL}$ amplitudes
$F^{ZZ}_{++00}$, $F^{WW}_{++00}$ vanish because of squark/quark cancelations;
while  the $HC_{LL}$  limits   in (\ref{HCLLasym}) continue to hold,
due to the vanishing of the asymptotic squark-loop contributions.\\

It is worth emphasizing at  this point that in both, MSSM and SM,
the asymptotic limits of the TT amplitudes depend solely on the gauge couplings;
while  the   LL ones  depend in addition on the top
and bottom masses\footnote{We neglect the quark masses of the first two generations in
this work.}.

It is moreover  impressive to observe the SM   $HV_{TT}$ and $HV_{LL}$ amplitudes  to
tend asymptotically to real  constants, which in MSSM
are exactly canceled  by opposite  squark contributions, due
to the beautiful HCns theorem.

Note that the behavior of the  $HV_{LL}$ amplitudes $F^{ZZ}_{++00}$, $F^{WW}_{++00}$ is
related by the equivalence theorem \cite{equiv1,equiv2, equiv3}, to the Goldstone amplitudes
discussed in \cite{ggHHVH}.
The $HV_{TT}$ results   though, are  new and indicate  strong violation
of HCns in  12 independent  SM amplitudes; compare  (\ref{F-ind-gammagamma}-\ref{F-ind-WW}).

The approach to the limiting values of the various $gg \to VV'$ amplitudes
mentioned above, appears   like $m/\sqrt{s}$ or $m^2/s$, multiplied by logarithms.
 Therefore,  the asymptotic limits    should be  reached rather early; i.e.
 as soon as the energy exceeds by  a few times  $\mw$ in SM, or  $M_{SUSY}$ in MSSM.\\

The limits  (\ref{Fasym1}), may be used to derive the asymptotic relation
\bqa
&& \frac{F_{\mu\mu'\tau\tau'}(gg\to \gamma \gamma)}{C_{\gamma \gamma q}}=
\frac{F_{\mu\mu'\tau\tau'}(gg\to \gamma Z)}{C_{\gamma Z q}} \nonumber \\
&& = \frac{F_{\mu\mu'\tau\tau'}(gg\to ZZ)}{C_{ZZ q}}=
\frac{F_{\mu\mu'\tau\tau'}(gg\to W^+W^-)}{C_{ WWq}} ~, \label{R8F}
\eqa
for  the $HC_{TT}$ amplitudes in MSSM, and all TT-amplitudes in SM.

In MSSM, where all other amplitudes vanish asymptotically,
(\ref{R8F}) may  be transformed to asymptotic  relations among unpolarized
cross sections, by subtracting the  LL-parts of the $gg\to VV'$-processes,
  using the $R_1,R_2, R_3$ relations of \cite{ggHHVH} and the equivalence theorem
  \cite{equiv1, equiv2, equiv3}. This way, in addition to the seven relations $(R_1,...~R_7)$
   derived in  \cite{ggHHVH}, we  obtain
\bqa
R_8 & \Rightarrow & \frac{\tilde \sigma(gg\to \gamma \gamma)}{C_{\gamma\gamma q}^2}=
\frac{\tilde \sigma(gg\to \gamma Z)}{C_{\gamma Z q}^2}
 =\frac{ \left[\tilde \sigma(gg\to ZZ)
-\left( \frac{R_{a1}}{R_{a5}}\right )^2 \tilde \sigma(gg\to h^0h^0) \right ]}{C_{ZZq}^2}
\nonumber \\[.1cm]
& = &\frac{\left[\tilde \sigma(gg\to W^+W^-)
-\left( \frac{R_{a1}}{R_{a5}}\right )^2 \tilde \sigma(gg\to h^0h^0)
-\left( \frac{R^J_{c1}}{R_{b2}}\right )^2 \tilde \sigma(gg\to Zh^0)  \right ]}{C_{WWq}^2}~~,
\label{R8sigma}
\eqa
where we have used the "dimensionless" unpolarized differential cross sections
\bqa
 \tilde \sigma(gg\to VV')  & \equiv & \frac{512 \pi}{\alpha^2 \alpha_s^2}\,
\frac{s^{3/2}}{p}\, {d\sigma (gg\to VV'; s) \over d\cos\theta}
=\sum_{\mu\mu'\tau\tau'}\frac{|F^{VV'}_{\mu\mu'\tau\tau'}|^2}{\alpha^2 \alpha_s^2}
~~~,  \label{sigma-tilde}
\eqa
and correspondingly for the $h^0h^0$ and $Zh^0$ production processes.
The constants needed in (\ref{R8sigma})  are
\bqa
&& R_{a1}  =  m^2_t+m^2_b ~~ ,  ~~~
R_{a5} =  {m^2_t\cos^2\alpha\over\sin^2\beta}+ {m^2_b\sin^2\alpha\over\cos^2\beta} ~~,
\nonumber \\
&& R_{b2}  =   {m^2_t\cos\alpha\over\sin\beta}+{m^2_b\sin\alpha\over\cos\beta}~~,~~
R^J_{c1}  =  m^2_t-m^2_b \simeq R_{a1} ~~, \label{R8couplings}
\eqa
 defined in  \cite{ggHHVH}, with $\alpha, \beta$ being the usual scalar sector angles
 in MSSM.  Eqs.(\ref{Casym2}) is also used in (\ref{R8sigma}).

 Relation  (\ref{R8sigma}) is an asymptotic MSSM prediction  involving
 measurable cross sections. It is on the same footing as the
 $R_1, ~R_2,~R_3,~R_4, ~R_5,$ relations listed  in (23-27) of \cite{ggHHVH}.
If MSSM is realized in Nature, their validity  is guaranteed,
provided the energy is sufficiently higher than the SUSY scale.\\

\section{Numerical results in MSSM and SM }

\begin{table}[h]
\begin{center}
{ Table 1: Asymptotic TT and LL  Helicity Amplitudes divided by $\alpha \alpha_s$, \\
 and asymptotic $\tilde \sigma(gg\to VV')$, in MSSM and SM, at $\theta=60^o$. }\\
  \vspace*{0.3cm}
\begin{small}
\begin{tabular}{||c|c|c||c|c|c||}
\hline \hline
\multicolumn{3}{||c||}{$gg \to \gamma \gamma $} & \multicolumn{3}{c||}{$gg \to \gamma Z $} \\
 \hline
 & MSSM & SM   &     & MSSM  & SM     \\ \hline
 $F_{++++}(\theta)$ & $-37. $  & $-26.$ &  $F_{++++}(\theta)$ & $-19.$  & $-13.5 $  \\
  $F_{+-+-}(\theta)$ & $-6.4 +i 29 $   & $-3.4 +i 20 $  &
      $F_{+-+-}(\theta)$ & $-3.3+i 15.$ & $-1.7+i 10.3 $   \\
   $F_{+--+}(\theta)$ & $-0.28+i  6.0 $ & $-0.14 +i 4.0 $
   &  $F_{+--+}(\theta)$ &  $-0.14+i 3.1 $ &  $-0.07+i 2.1 $  \\
  $F_{++--}(\theta)$ & 0 & $6.7$  &  $F_{++--}(\theta)$ & 0 & $3.4 $  \\
 $F_{+++-}(\theta)$ & 0 & $6.7$ &  $F_{+++-}(\theta)$ & 0 & $3.4 $ \\
   & &  &  $F_{++-+}(\theta)$ & 0 & $3.4 $ \\
 & &    &  $F_{+-++}(\theta)$ & 0 & $3.4$   \\
 $\tilde \sigma(gg\to \gamma \gamma)$ &  $4567$ & $ 2654$  &
 $\tilde \sigma(gg\to \gamma Z)$ & $1220 $  & $709 $  \\
\hline  \hline
\multicolumn{3}{||c||}{$gg \to ZZ $} & \multicolumn{3}{c||}{$gg \to W^+W^- $} \\
 \hline
 & MSSM  & SM  &     & MSSM  & SM     \\ \hline
 $F_{++++}(\theta)$ & $-61$ & $-43$  &  $F_{++++}(\theta)$ & $-72.$ & $-51.$  \\
  $F_{+-+-}(\theta)$ & $-10.6+i  48.1 $ & $-5.6+i 33. $
  &   $F_{+-+-}(\theta)$ & $-12.+i 56. $  & $-6.5+i  39 $   \\
   $F_{+--+}(\theta)$ & $-0.46+i 10.$ & $ -0.23+i 6.7 $
   &  $F_{+--+}(\theta)$ & $-0.5+i 12. $ &  $-0.27+i 7.8 $   \\
  $F_{++--}(\theta)$ & 0 & $ 11. $   &  $F_{++--}(\theta)$ & 0 &  $12.9 $  \\
 $F_{+++-}(\theta)$ & 0 & $11.$ &  $F_{+++-}(\theta)$ & 0 & $12.9$ \\
$F_{+-++}(\theta)$  & 0 &  $11.$ &  $F_{+-++}(\theta)$ & 0 & $12.9$  \\
$F_{+-00}(\theta)$  & $-4.6 +i 43.$   & $-4.6 +i 43. $
&   $F_{+-00}(\theta)$ & $-2.6+i 56.$      & $-2.6+i 56. $ \\
 $F_{++00}(\theta)$ & 0 &  $-20.5$  &   $F_{++00}(\theta)$ & 0 &  $-20.5 $  \\
 $\tilde \sigma(gg\to ZZ)$ &  $ 16226 $ & $11820$&
 $\tilde \sigma(gg\to W^+W^-)$ & $21232$  & $14873$  \\
  \hline \hline
\end{tabular}
 \end{small}
\end{center}
\end{table}

The asymptotic values of all TT and LL amplitudes for
$gg\to VV'$  in SM and MSSM,  only depend on the gauge
couplings and the top and bottom masses; see Sect.3. The corresponding
numerical results  are given
in Table 1. The  longitudinal-transverse ($HV_{TL}$) asymptotic amplitudes
are not presented, since they all vanish, for both MSSM and SM.

As seen from this Table, all  HV amplitudes indeed asymptotically vanish in MSSM,
in agreement with HCns.

On the contrary  for SM, the full number of the 12 independent  $HV_{TT}$ amplitudes
and the two $HV_{LL}$ ones, acquire  non-vanishing
asymptotic values,  comparable to those of the HC amplitudes.
Moreover, these SM asymptotic amplitudes only depend on the process;  they
are independent of the scattering angle and the specific helicities;
see (\ref{Casym2}, \ref{ASHV}, \ref{HVLLasym}).

A further conclusion we can draw is that, contrary to the situation
in  $\gamma \gamma \to \gamma \gamma, ~ \gamma Z, ZZ$,
where HCns is approximately satisfied in SM \cite{gamgamgamgam, gamgamgamZ, gamgamZZ};
it is strongly violated
in the $gg\to VV'$ ~SM amplitudes.

In the same Table,
the asymptotic values for the "dimensionless" unpolarized differential
cross sections defined in
(\ref{sigma-tilde}) are also given. \\

The  analytic calculation of the diagrams in Fig.\ref{diagrams-fig} is used to
construct the  ggvvcode Fortran  code computing  the helicity amplitudes
of the processes (\ref{processes}) in SM and MSSM, as functions of the c.m. energy in TeV
and the c.m. angle in radians. The amplitude conventions
are given immediately after (\ref{Fdef}),
 while its overall normalization  is  fixed by (\ref{sigma-tilde}).
 A factor $\alpha \alpha_s$ has been removed from the amplitudes, in the code-output.
All input  parameters are taken real  and  at the electroweak scale.
 The code, accompanied by a Readme file  explaining
its compilation and use, may be downloaded from \cite{code}.
The asymptotic limits presented in Table 1 agree with the results of the code.\\

\begin{figure}[p]
\vspace*{-1cm}
\[
\hspace{-0.5cm}
\epsfig{file=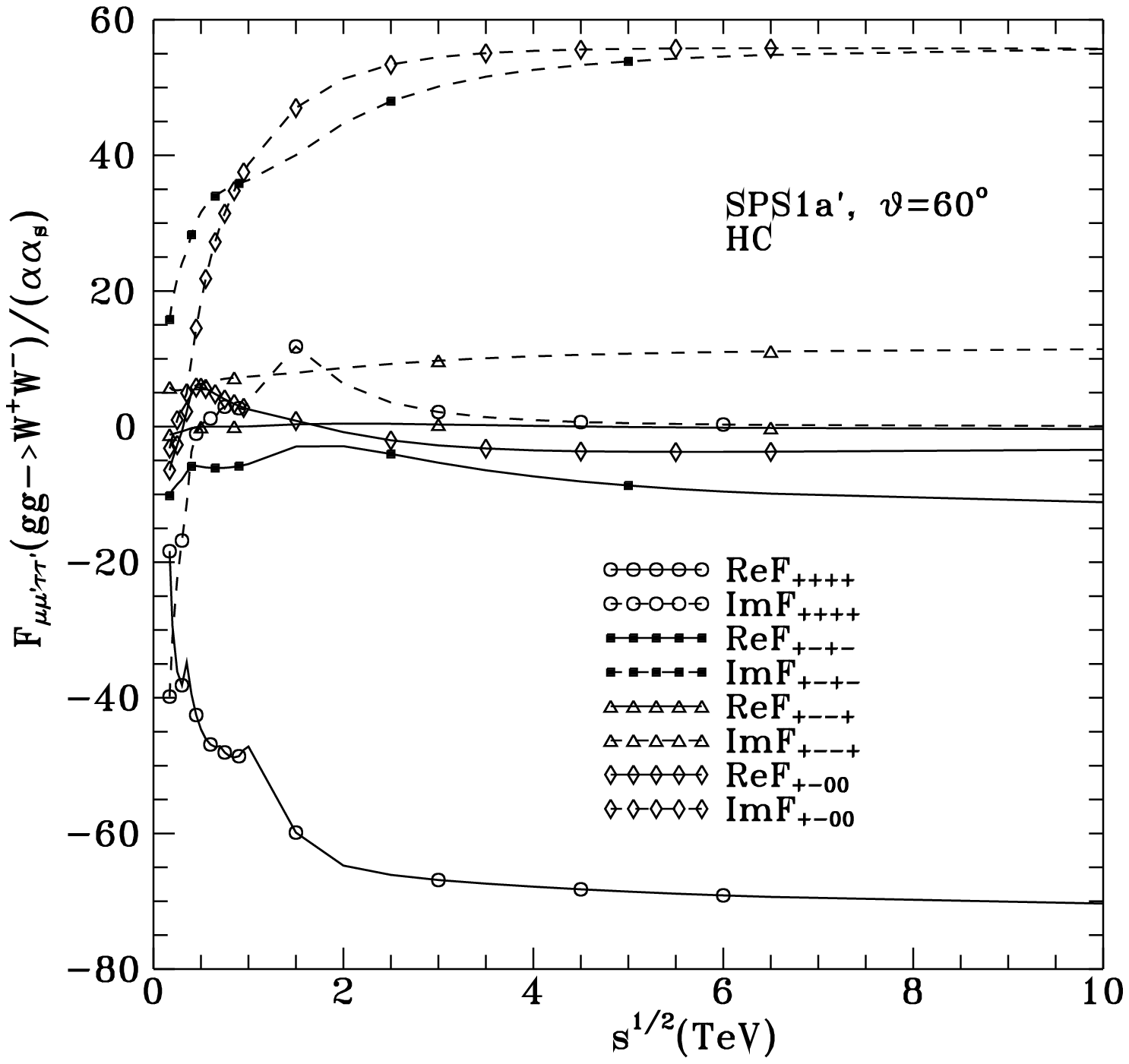, height=6.cm}\hspace{1.cm}
\epsfig{file=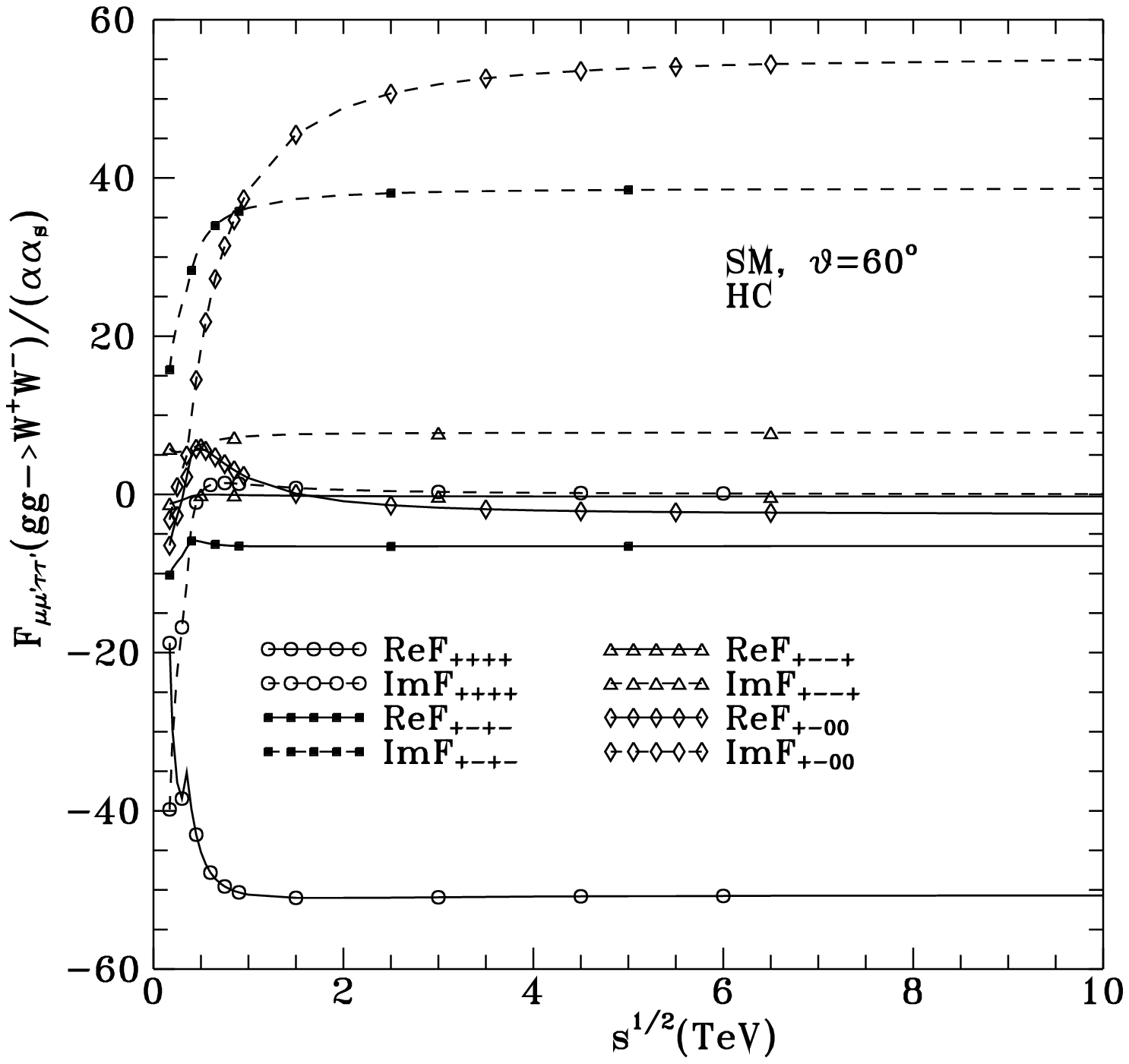,height=6.cm}
\]
\[
\hspace{-0.5cm}
\epsfig{file=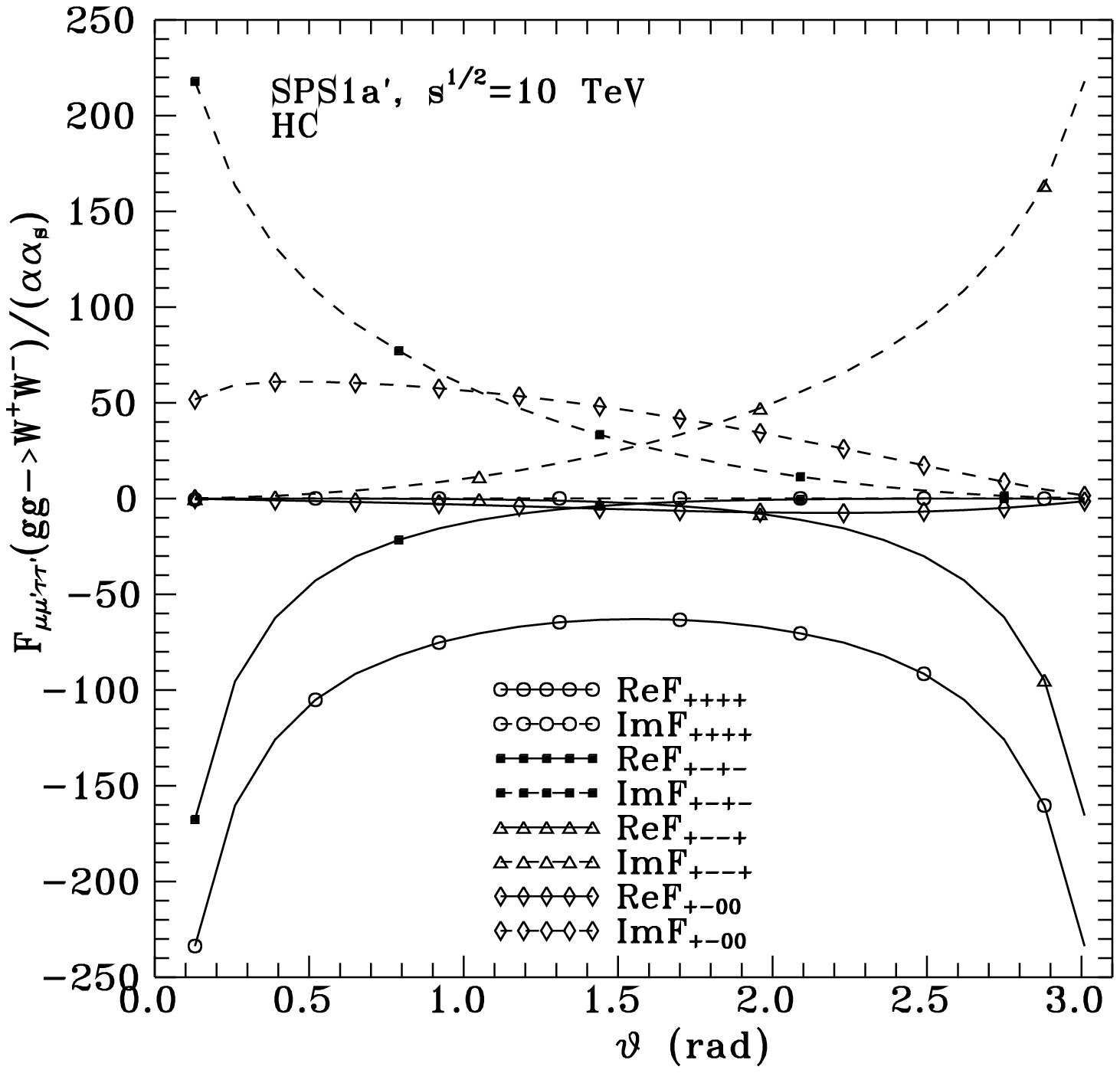, height=6.cm}\hspace{1.cm}
\epsfig{file=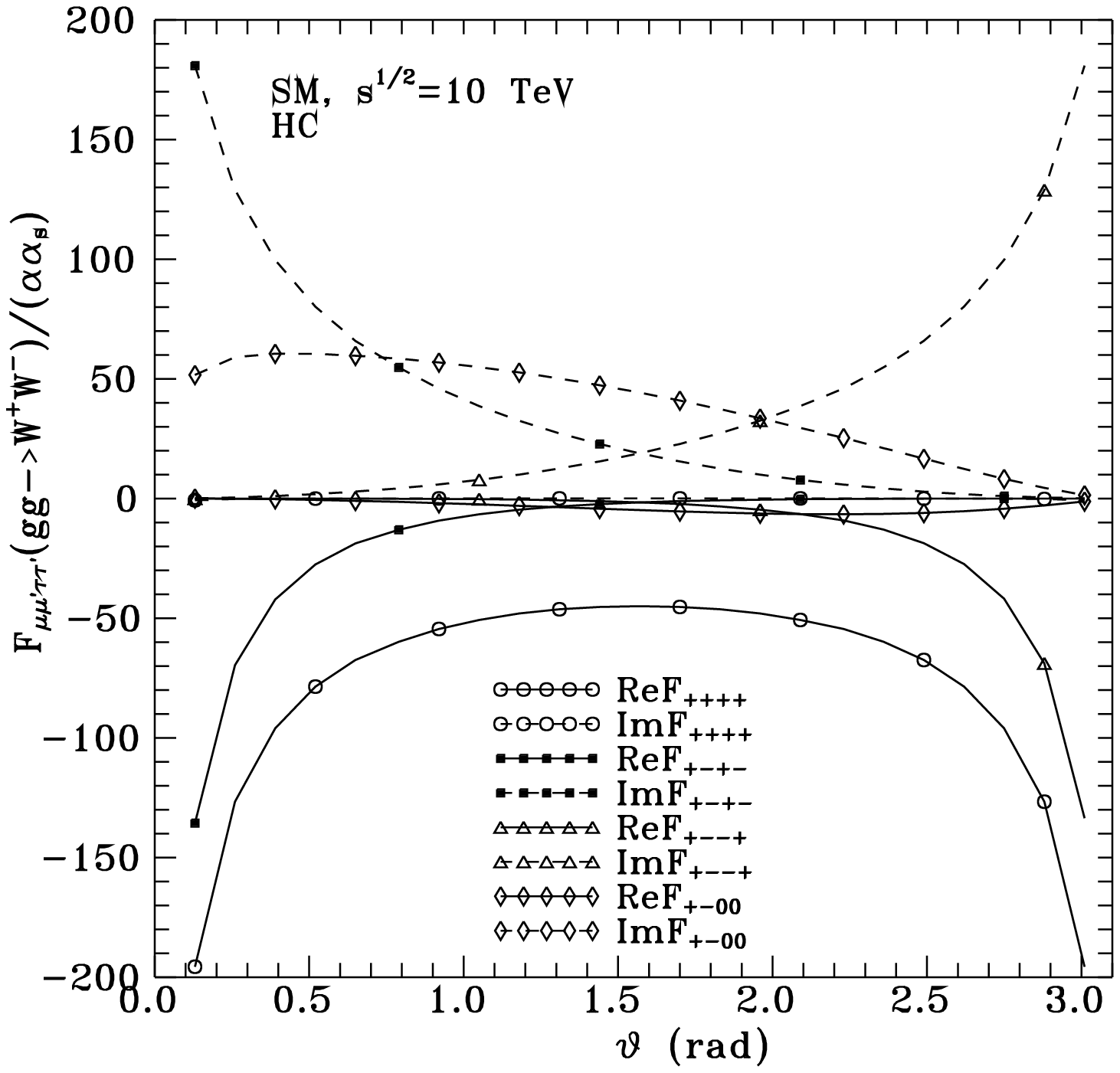,height=6.cm}
\]
\caption[1]{HC amplitudes for $gg\to W^+W^- $ in $SPS1a'$ and SM.
Upper panels give the energy dependencies at a fixed
c.m. angle $\theta=60^o$; while lower panels give the angular
distributions for a c.m. energy $\sqrt{s}=10~TeV$. Left panels always
 give  the    MSSM ($SPS1a'$) result \cite{SPA1},
while  right panels the SM one. }
\label{HC-ggWW-amp-fig}
\end{figure}

\begin{figure}[p]
\vspace*{-1cm}
\[
\hspace{-0.5cm}
\epsfig{file=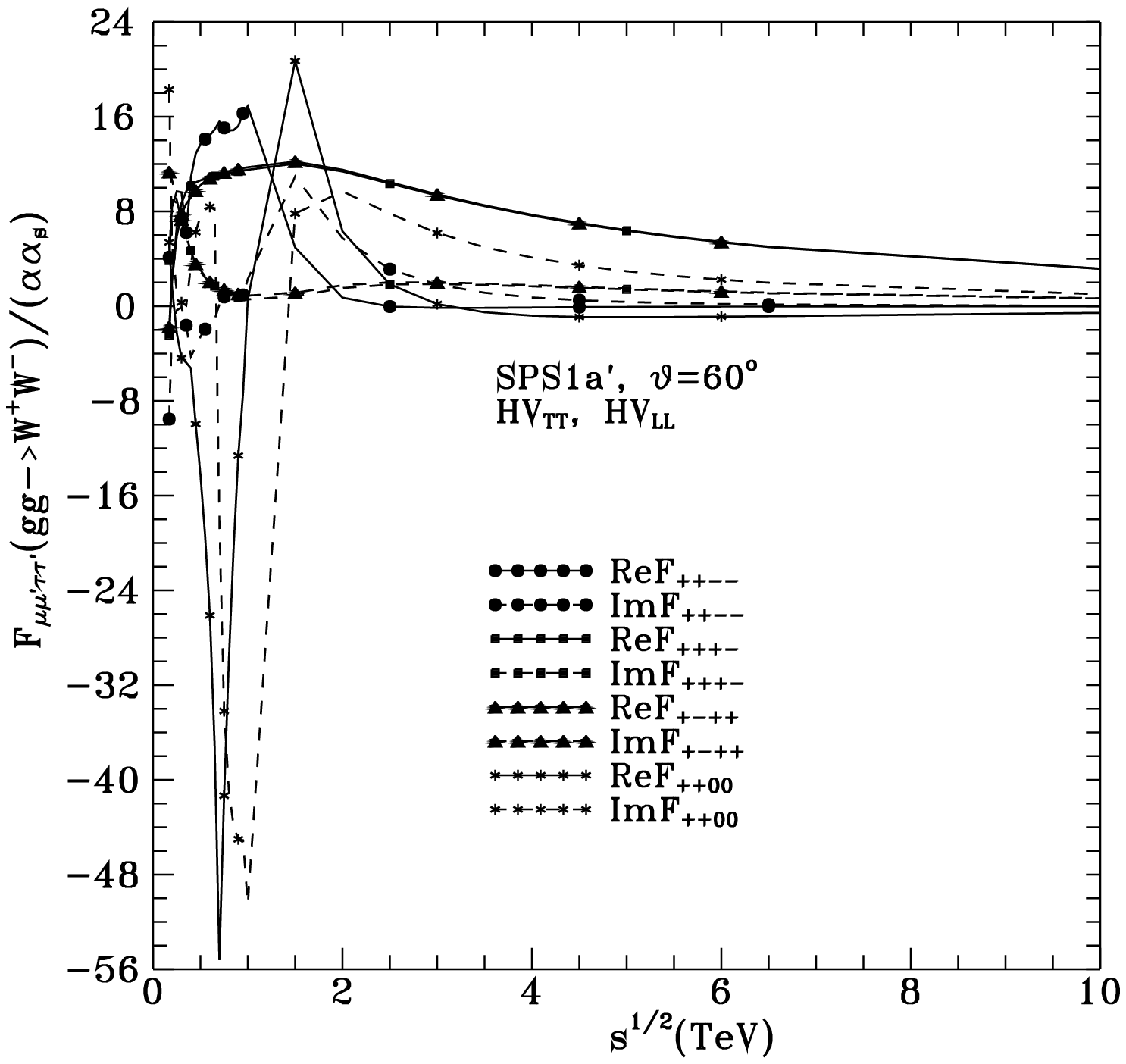, height=6.cm}\hspace{1.cm}
\epsfig{file=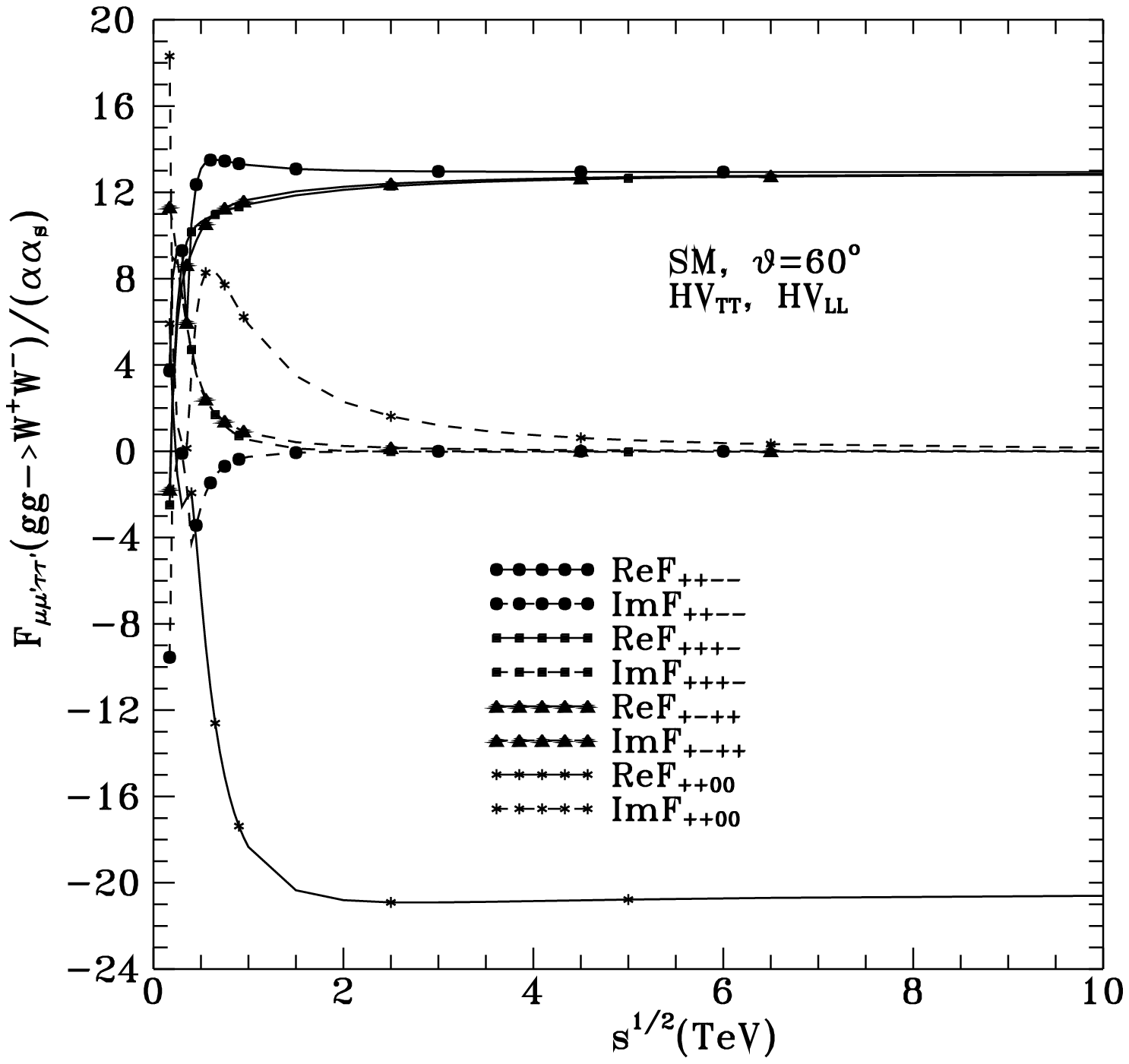,height=6.cm}
\]
\[
\hspace{-0.5cm}
\epsfig{file=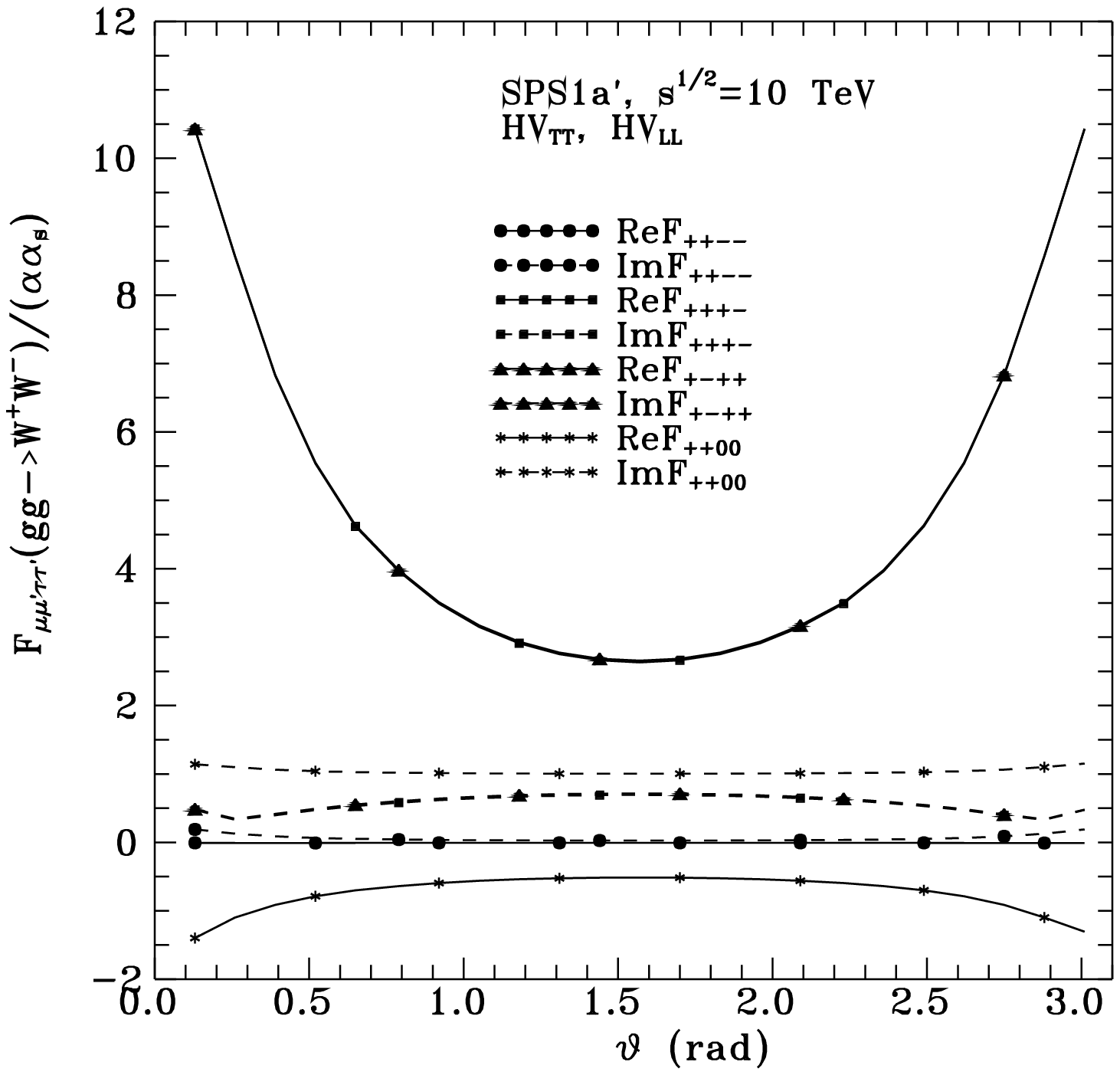, height=6.cm}\hspace{1.cm}
\epsfig{file=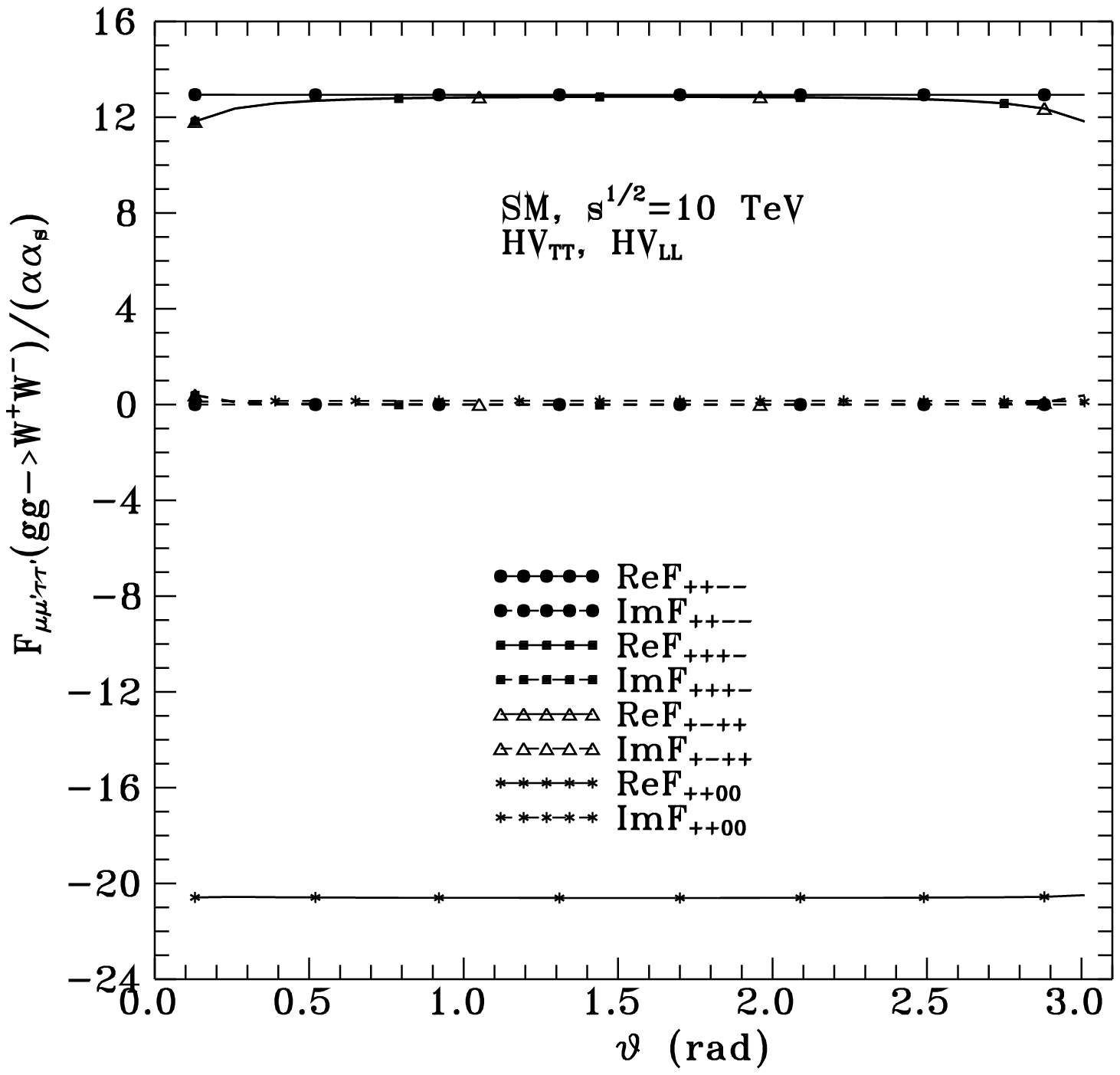,height=6.cm}
\]
\caption[1]{$HV_{TT}$ and $HV_{LL}$ amplitudes for $gg\to W^+W^- $ in $SPS1a'$ and SM,
as in Fig.\ref{HC-ggWW-amp-fig}.  }
\label{HV-TT-LL-ggWW-amp-fig}
\end{figure}

\begin{figure}[p]
\vspace*{-1cm}
\[
\hspace{-0.5cm}
\epsfig{file=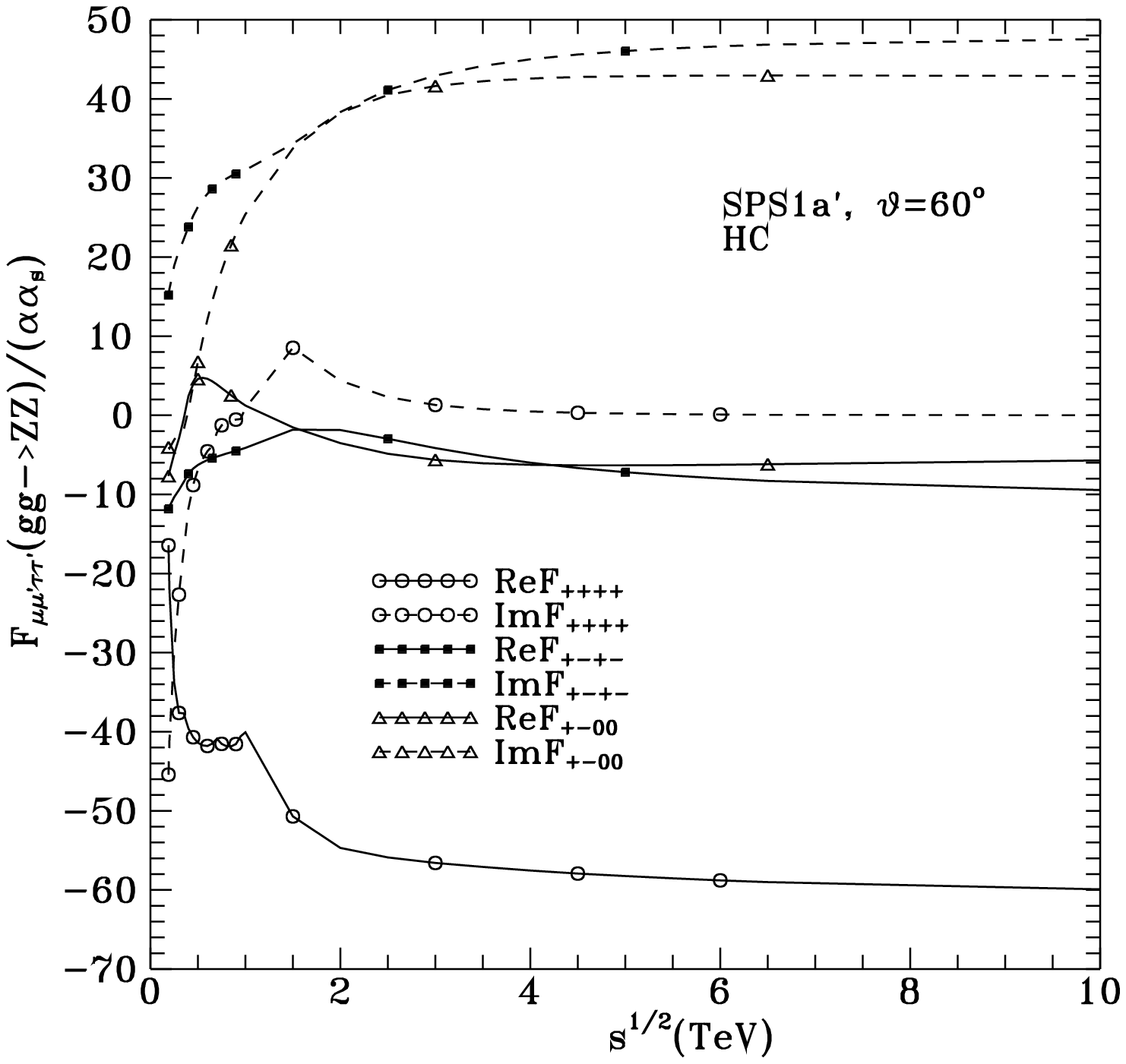, height=6.cm}\hspace{1.cm}
\epsfig{file=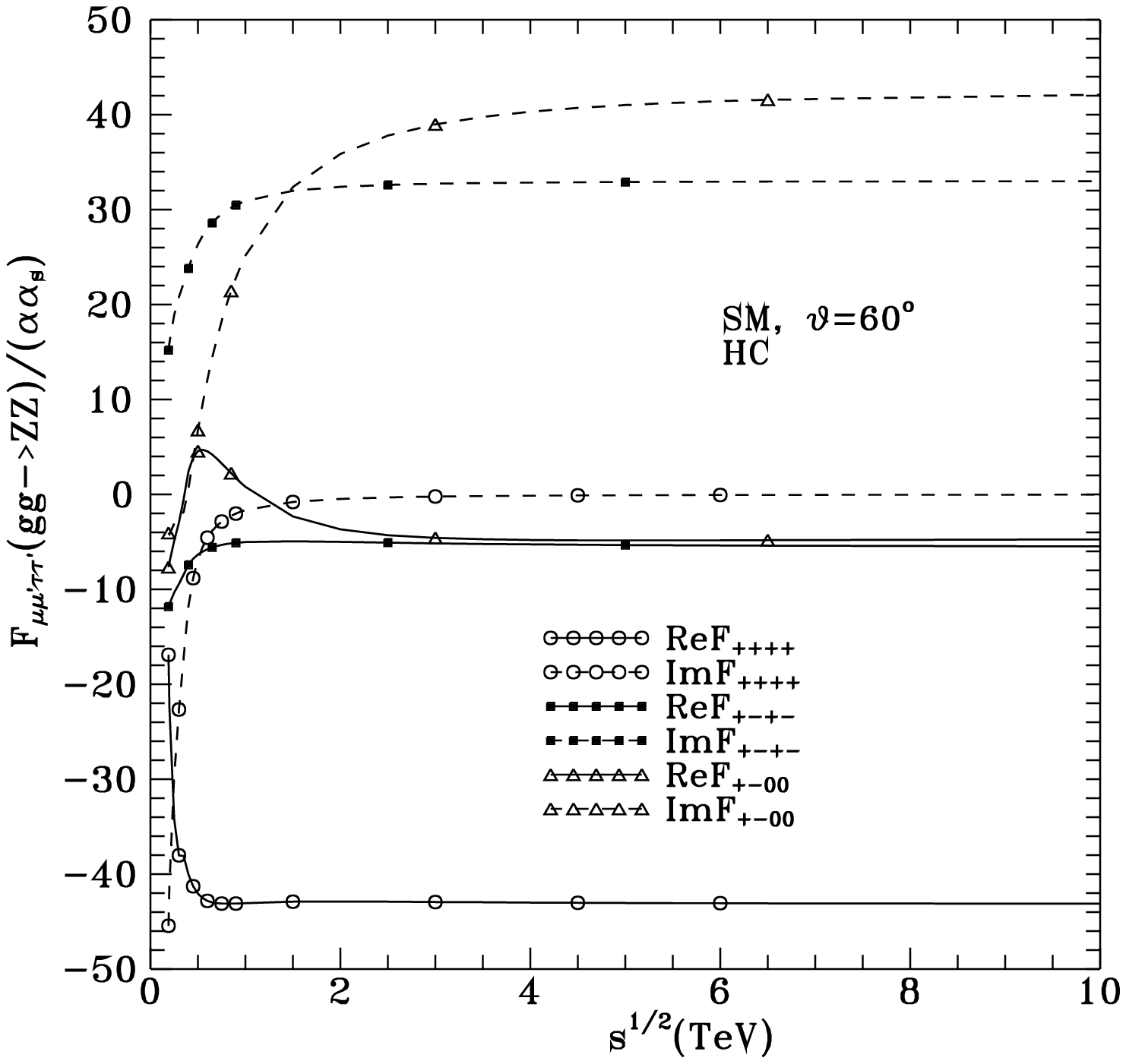,height=6.cm}
\]
\[
\hspace{-0.5cm}
\epsfig{file=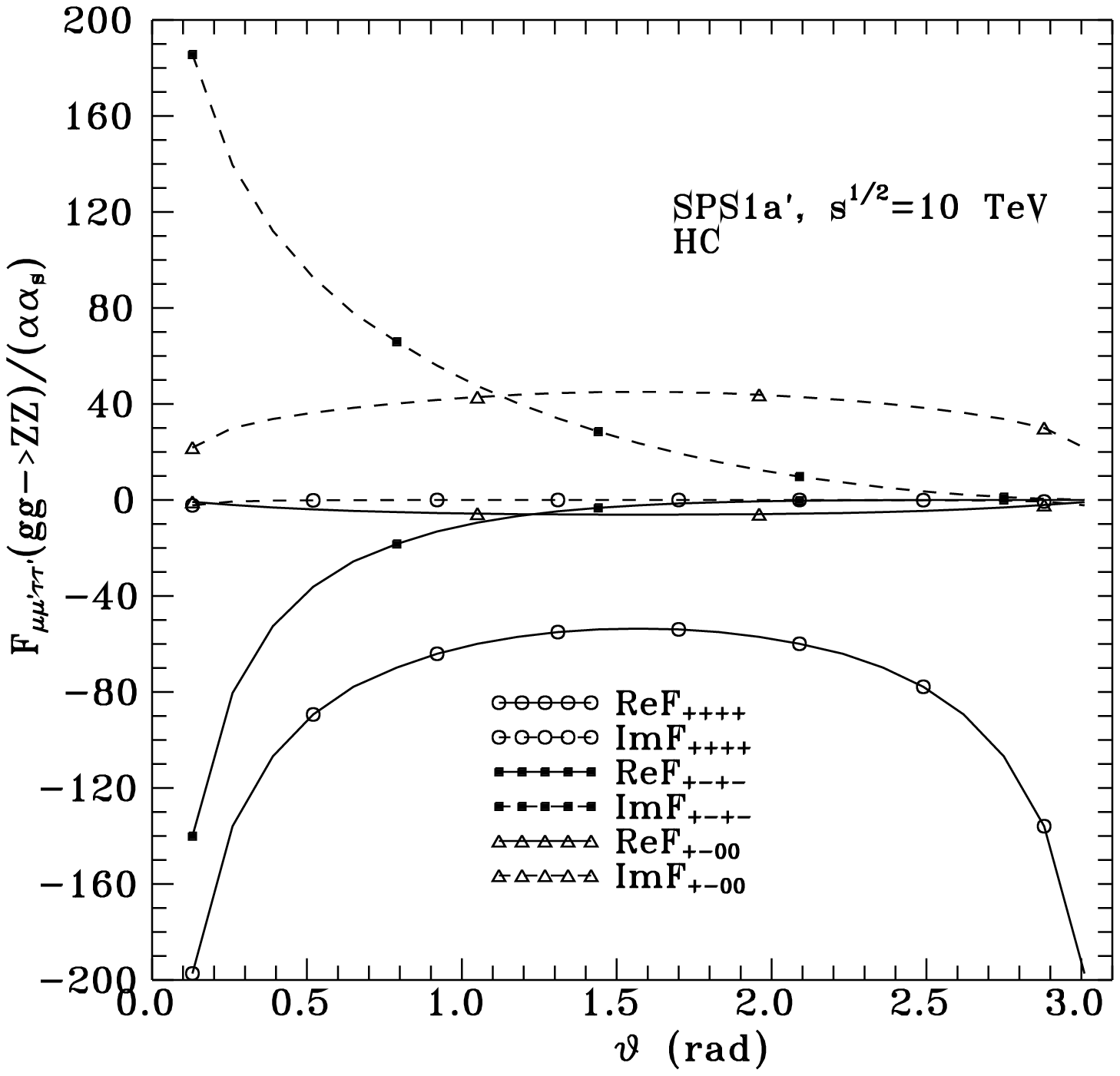, height=6.cm}\hspace{1.cm}
\epsfig{file=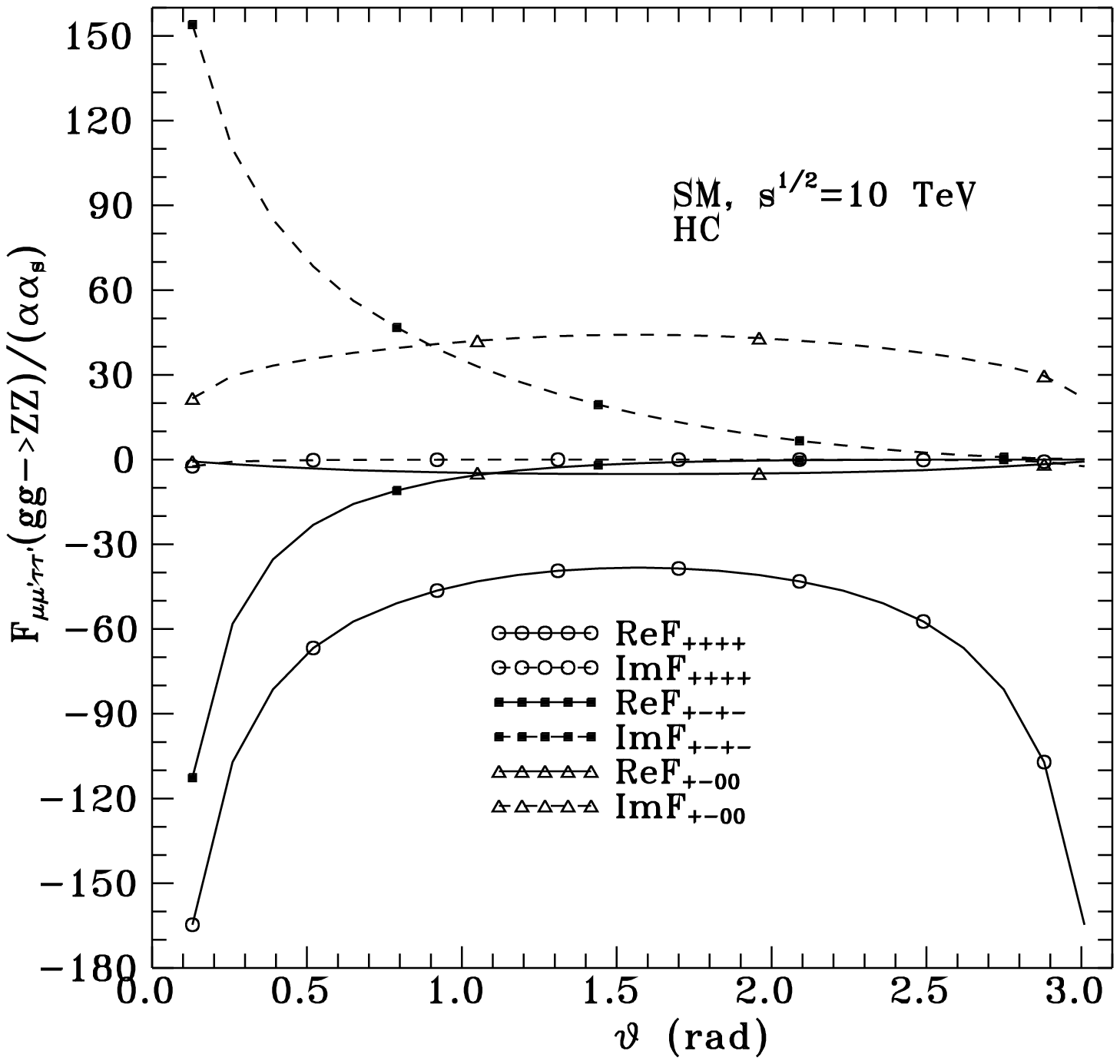,height=6.cm}
\]
\caption[1]{HC amplitudes for $gg\to Z Z $ in $SPS1a'$ and SM,
as in Fig.\ref{HC-ggWW-amp-fig}.  }
\label{HC-ggZZ-amp-fig}
\end{figure}

\begin{figure}[p]
\vspace*{-1cm}
\[
\hspace{-0.5cm}
\epsfig{file=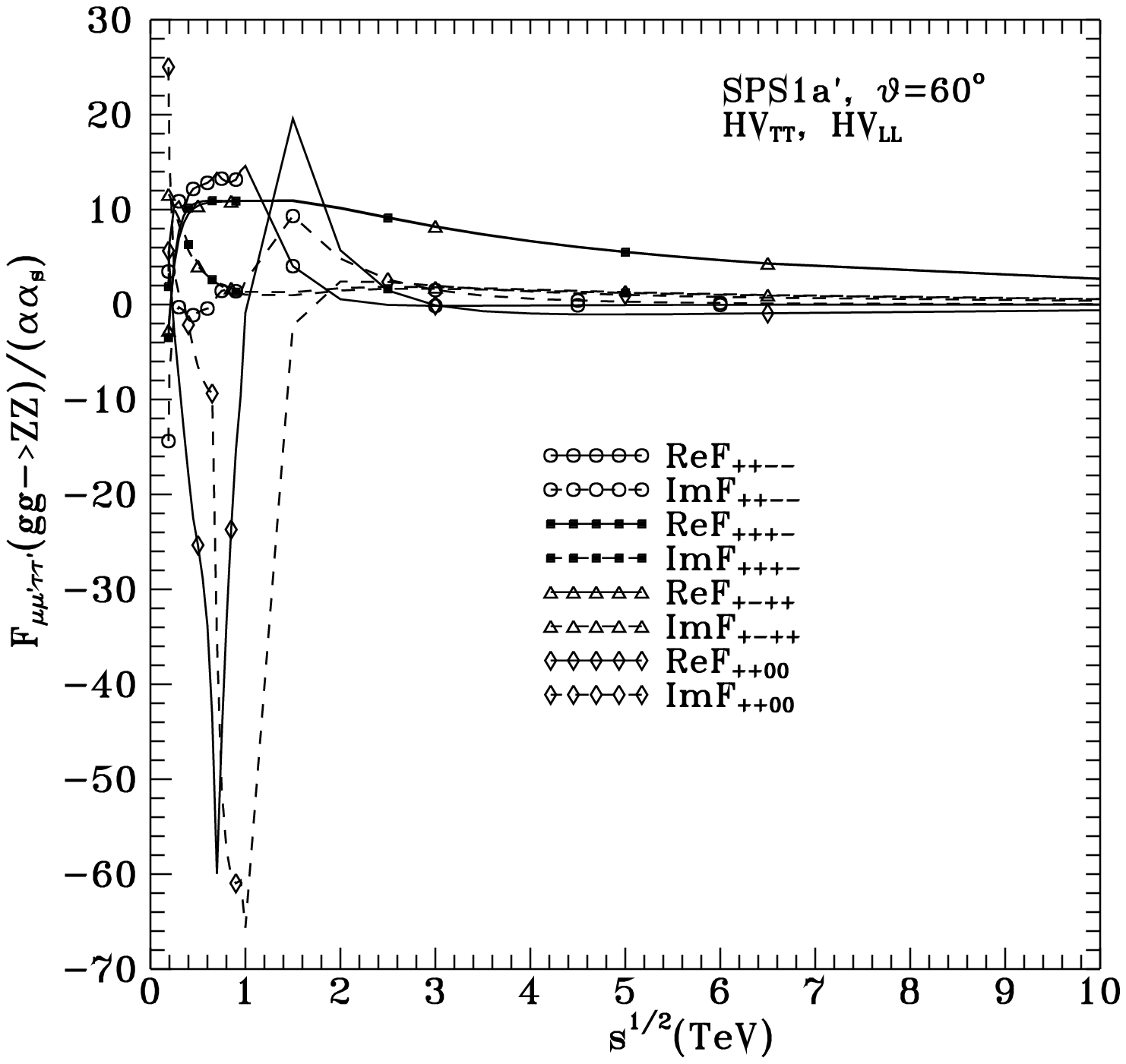, height=6.cm}\hspace{1.cm}
\epsfig{file=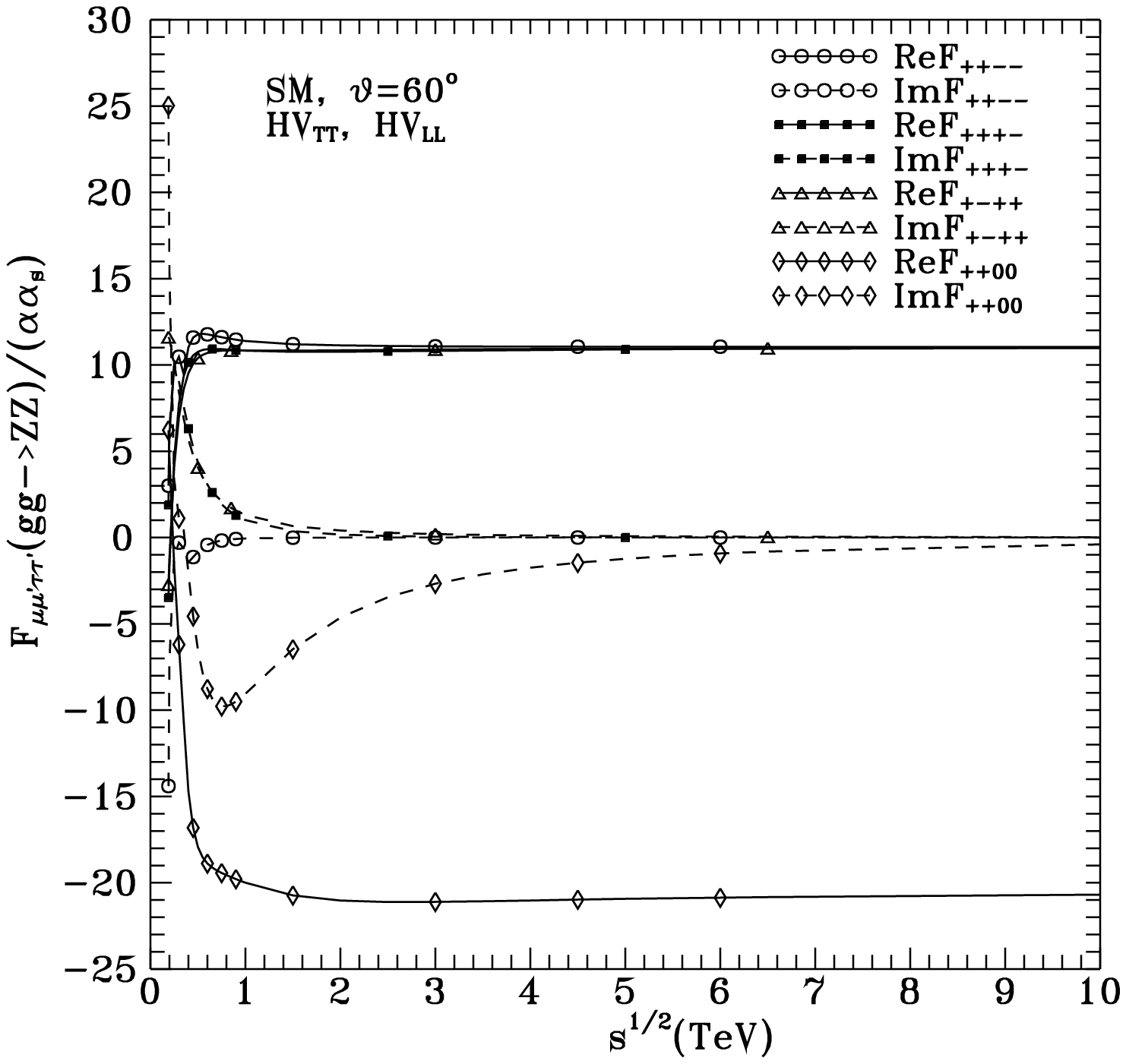,height=6.cm}
\]
\[
\hspace{-0.5cm}
\epsfig{file=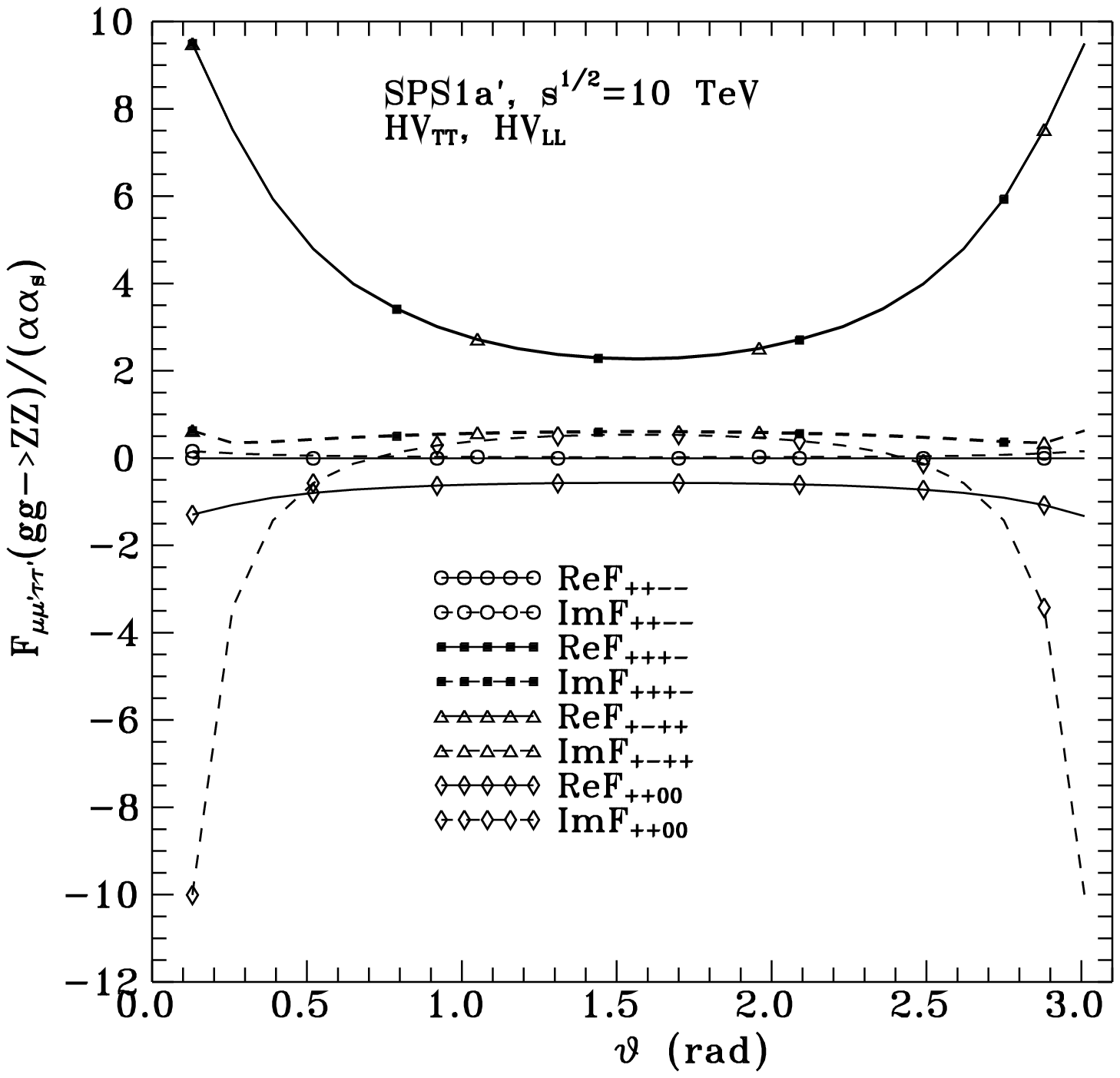, height=6.cm}\hspace{1.cm}
\epsfig{file=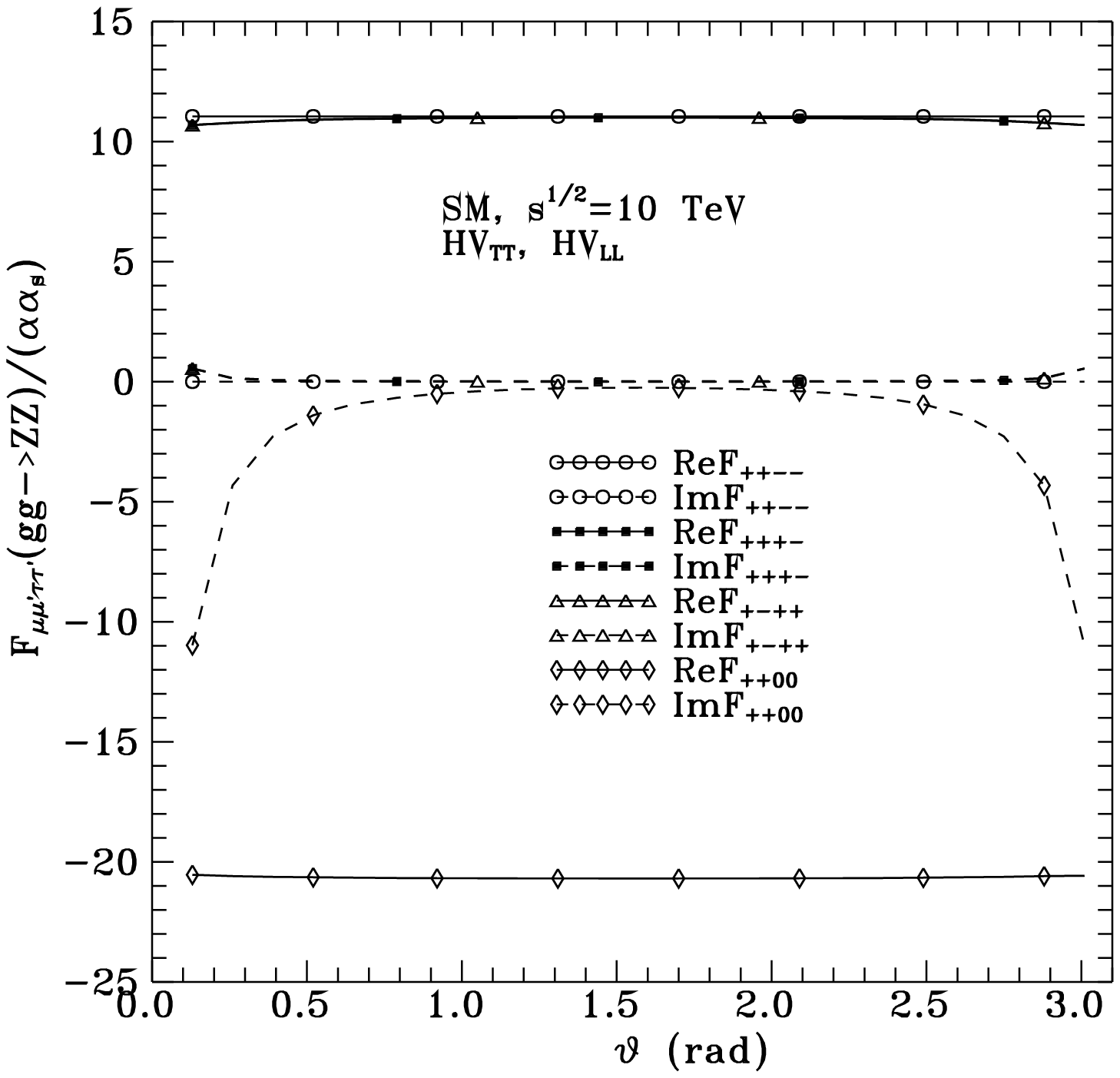,height=6.cm}
\]
\caption[1]{$HV_{TT}$ and $HV_{LL}$ amplitudes for $gg\to ZZ $ in $SPS1a'$ and SM,
as in Fig.\ref{HC-ggWW-amp-fig}. }
\label{HV-TT-LL-ggZZ-amp-fig}
\end{figure}

\begin{figure}[p]
\vspace*{-1cm}
\[
\hspace{-0.5cm}
\epsfig{file=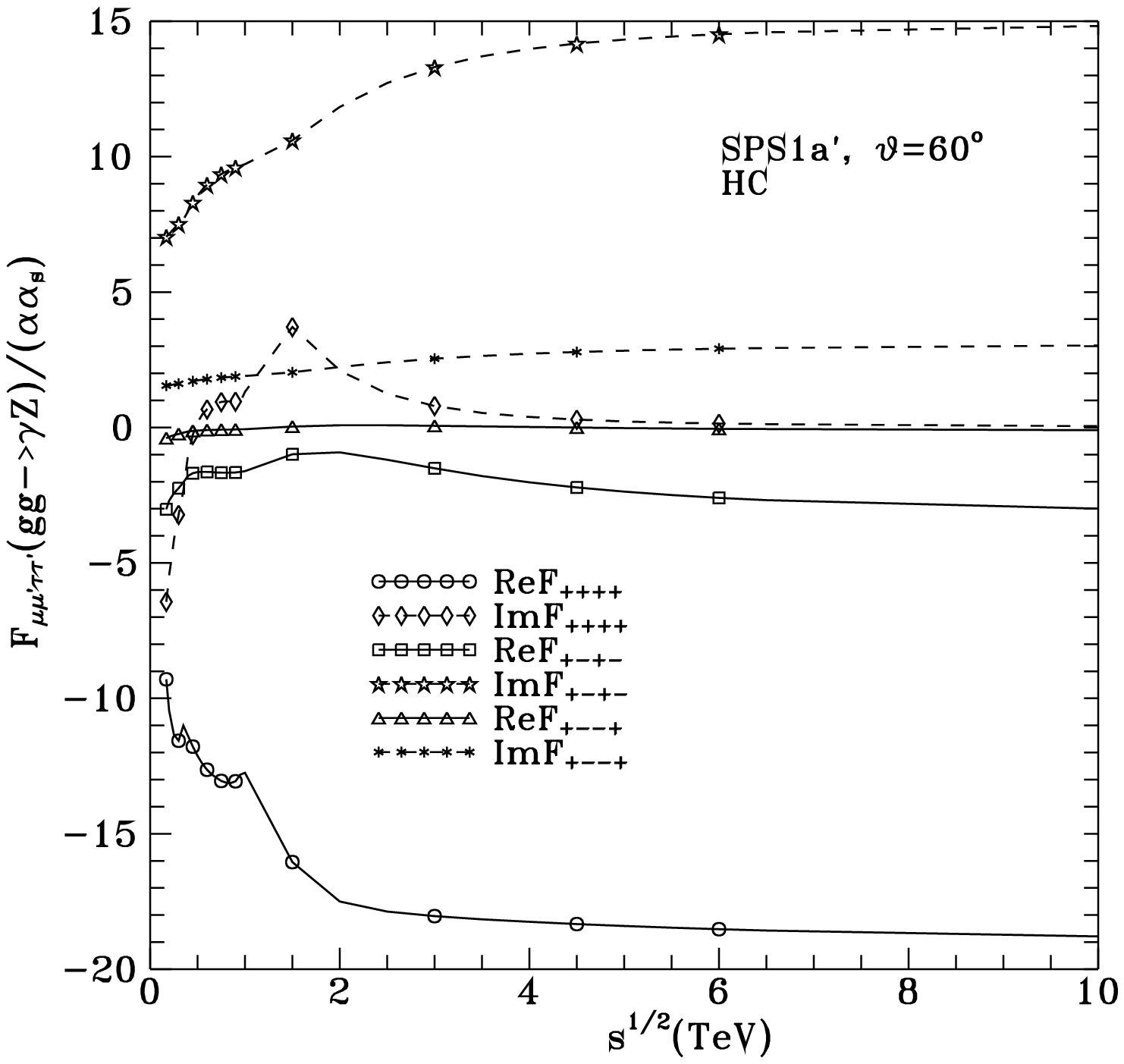, height=6.cm}\hspace{1.cm}
\epsfig{file=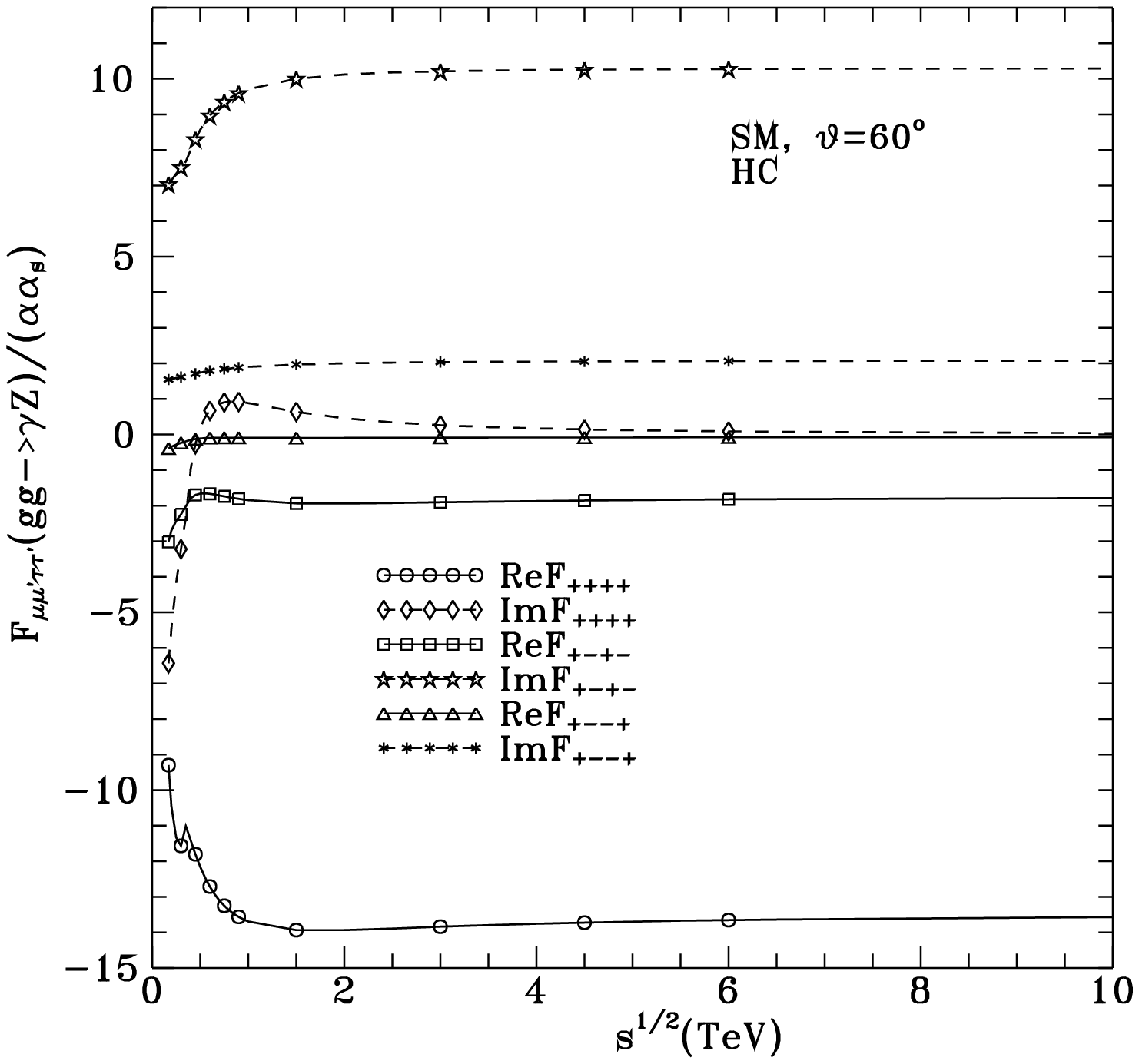,height=6.cm}
\]
\[
\hspace{-0.5cm}
\epsfig{file=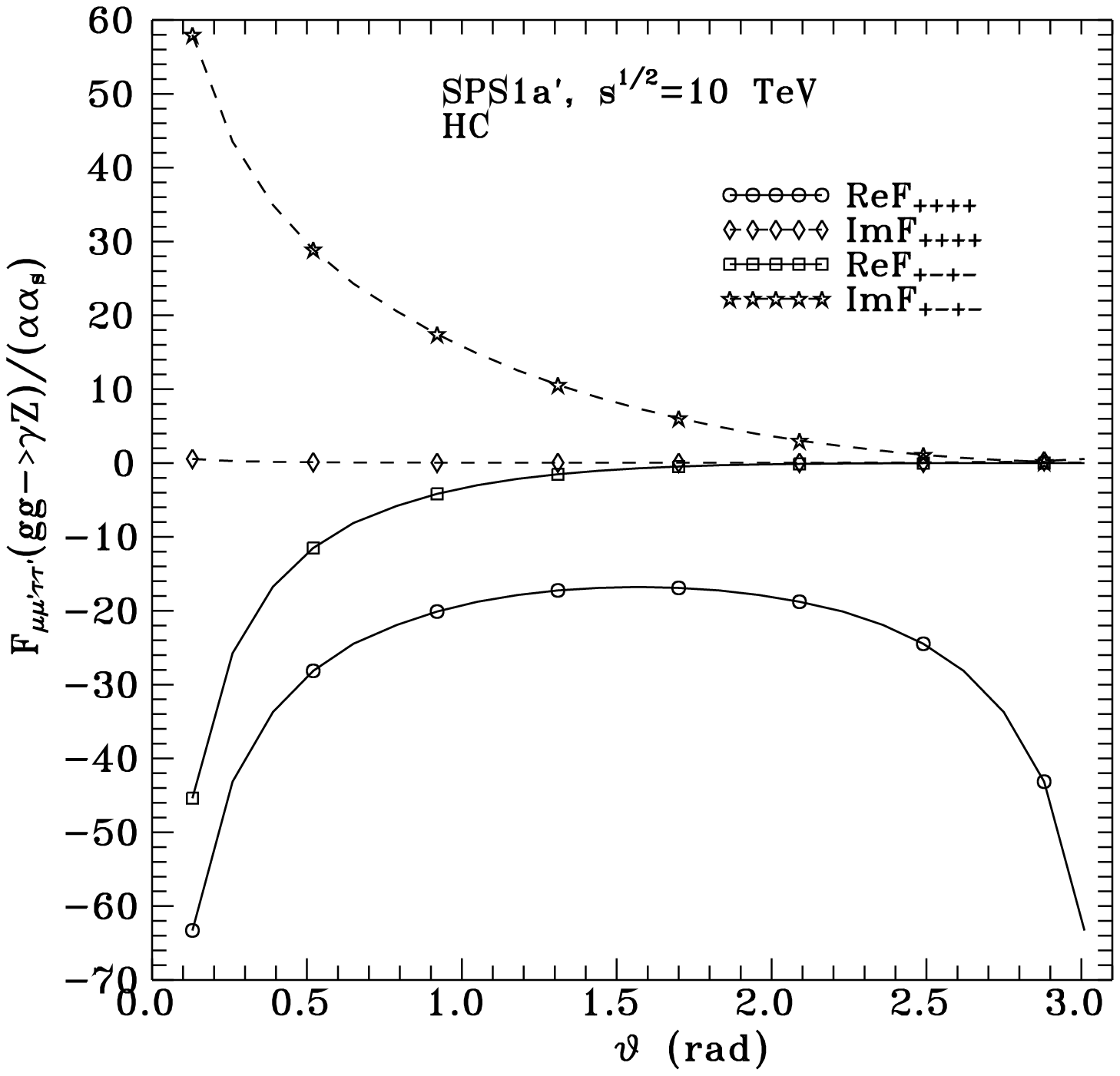, height=6.cm}\hspace{1.cm}
\epsfig{file=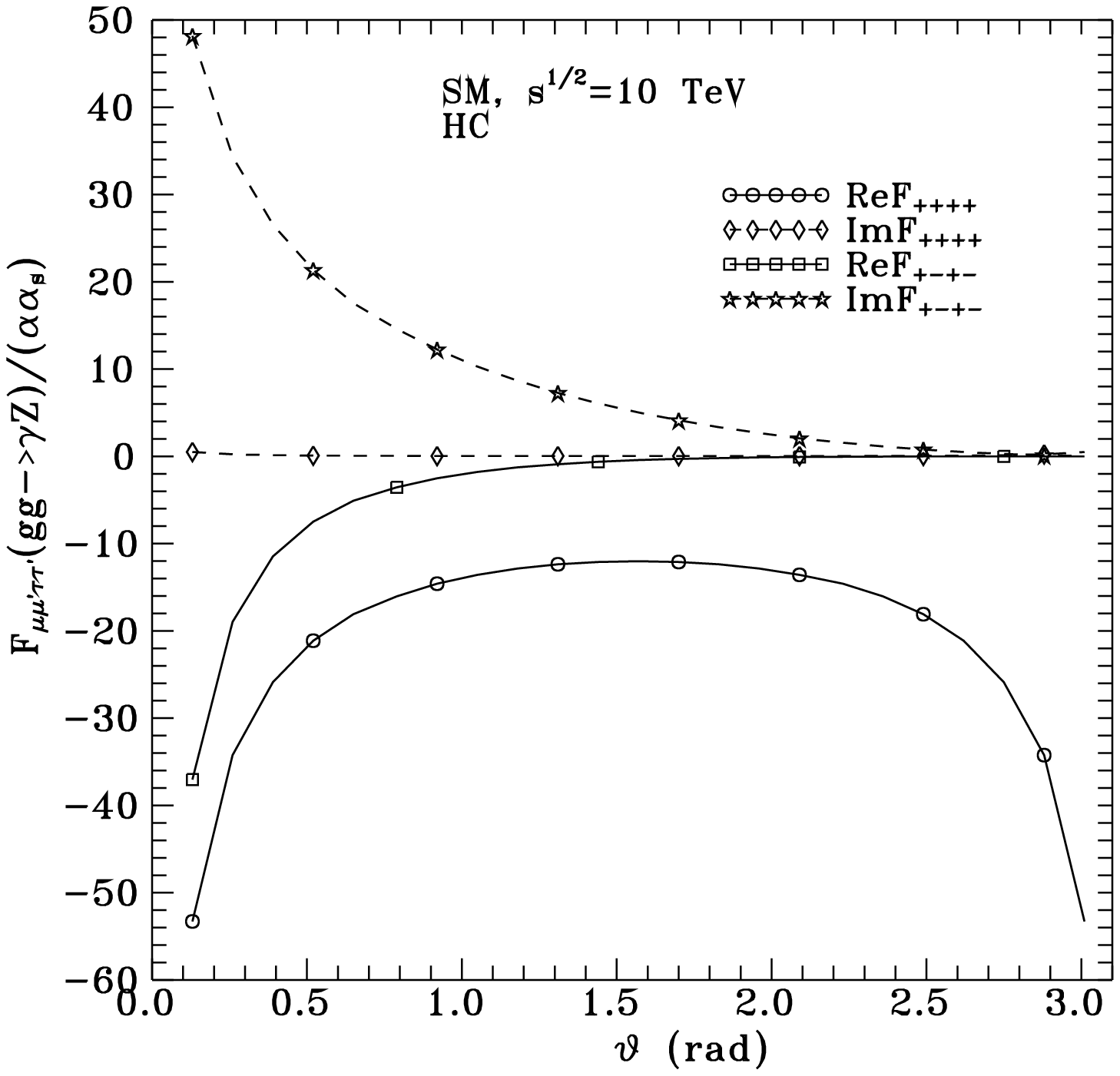,height=6.cm}
\]
\caption[1]{HC amplitudes for $gg\to \gamma Z $ in $SPS1a'$ and SM,
as in Fig.\ref{HC-ggWW-amp-fig}.  }
\label{HC-gggZ-amp-fig}
\end{figure}

\begin{figure}[p]
\vspace*{-1cm}
\[
\hspace{-0.5cm}
\epsfig{file=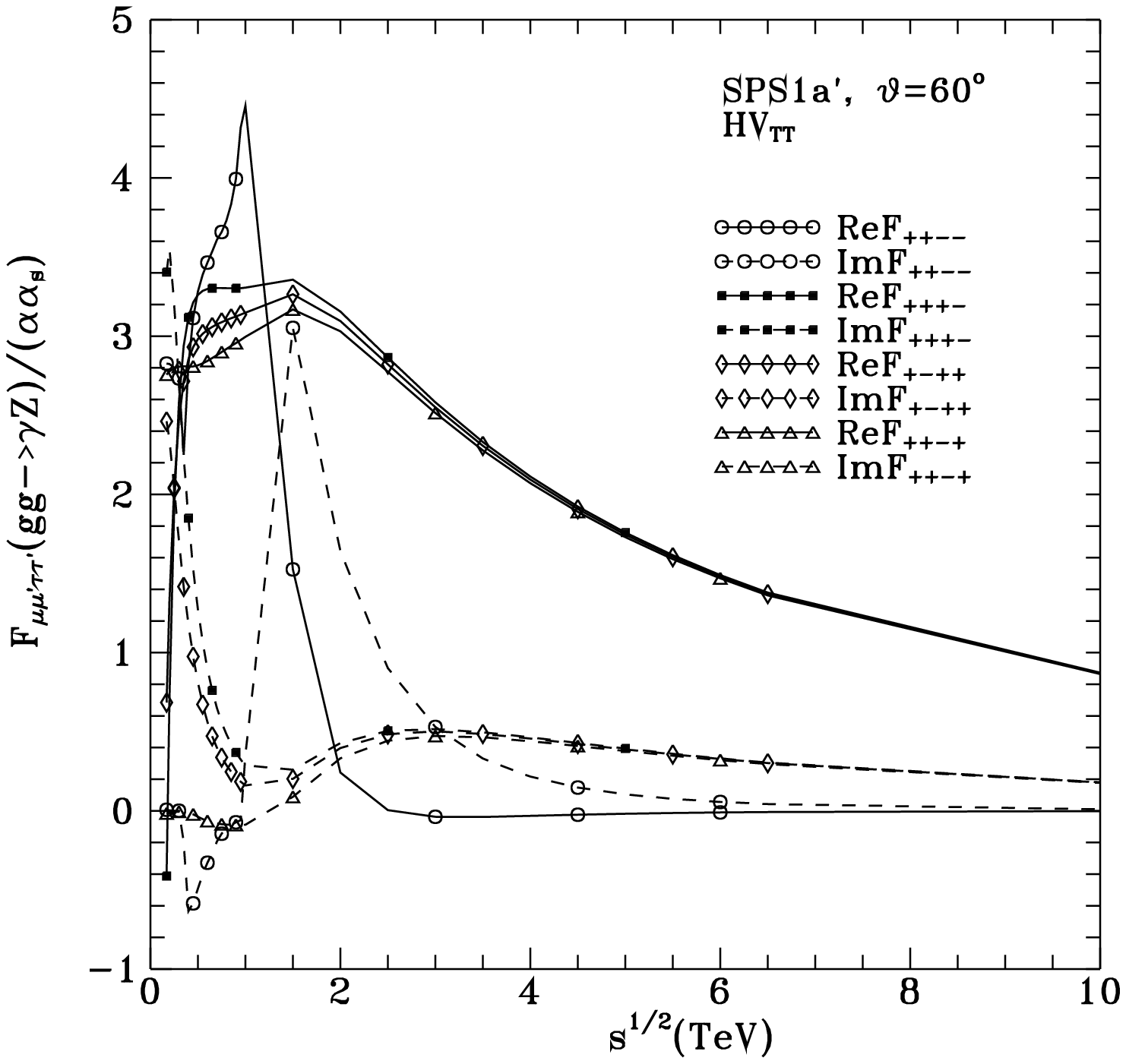, height=6.cm}\hspace{1.cm}
\epsfig{file=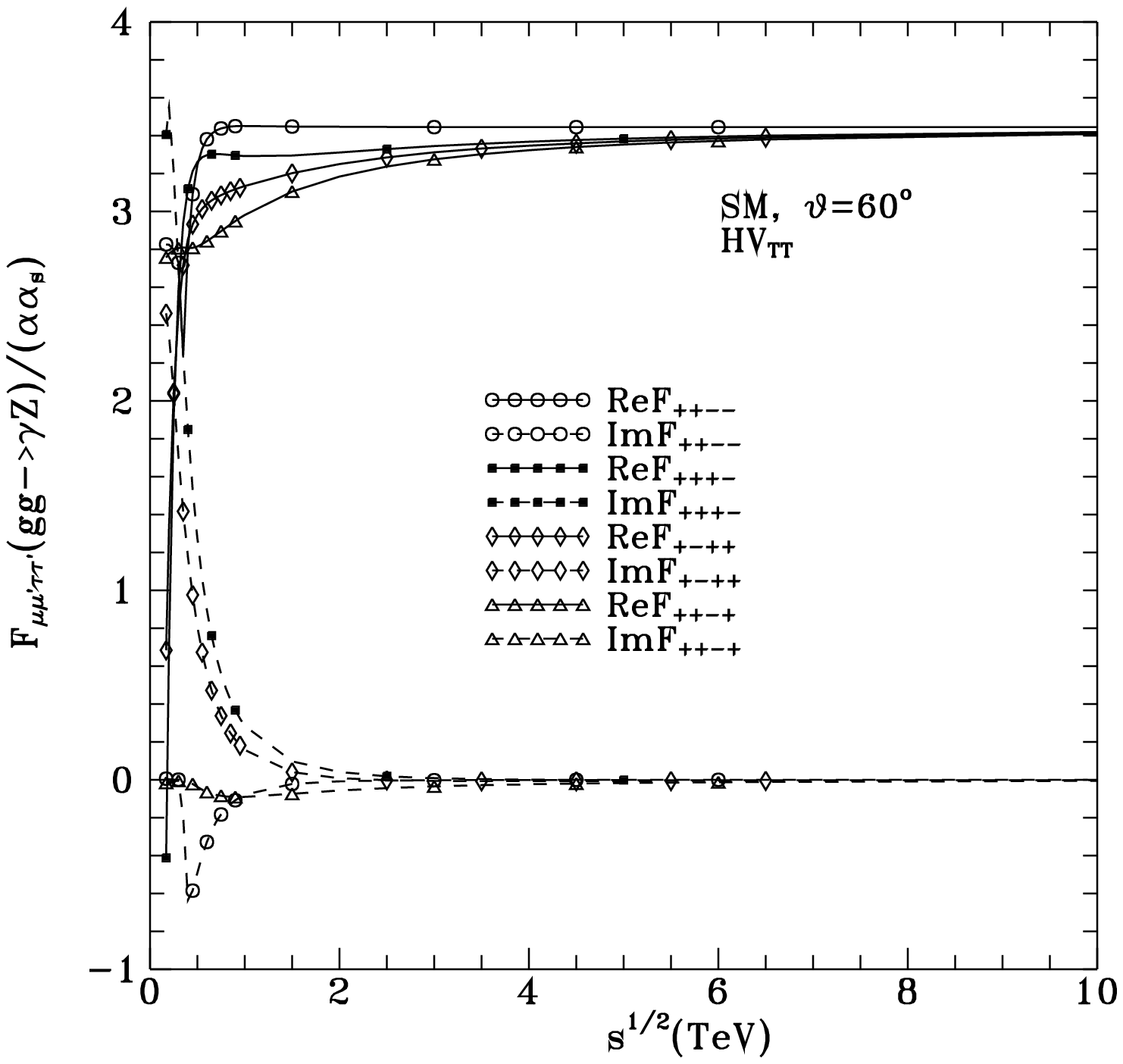,height=6.cm}
\]
\[
\hspace{-0.5cm}
\epsfig{file=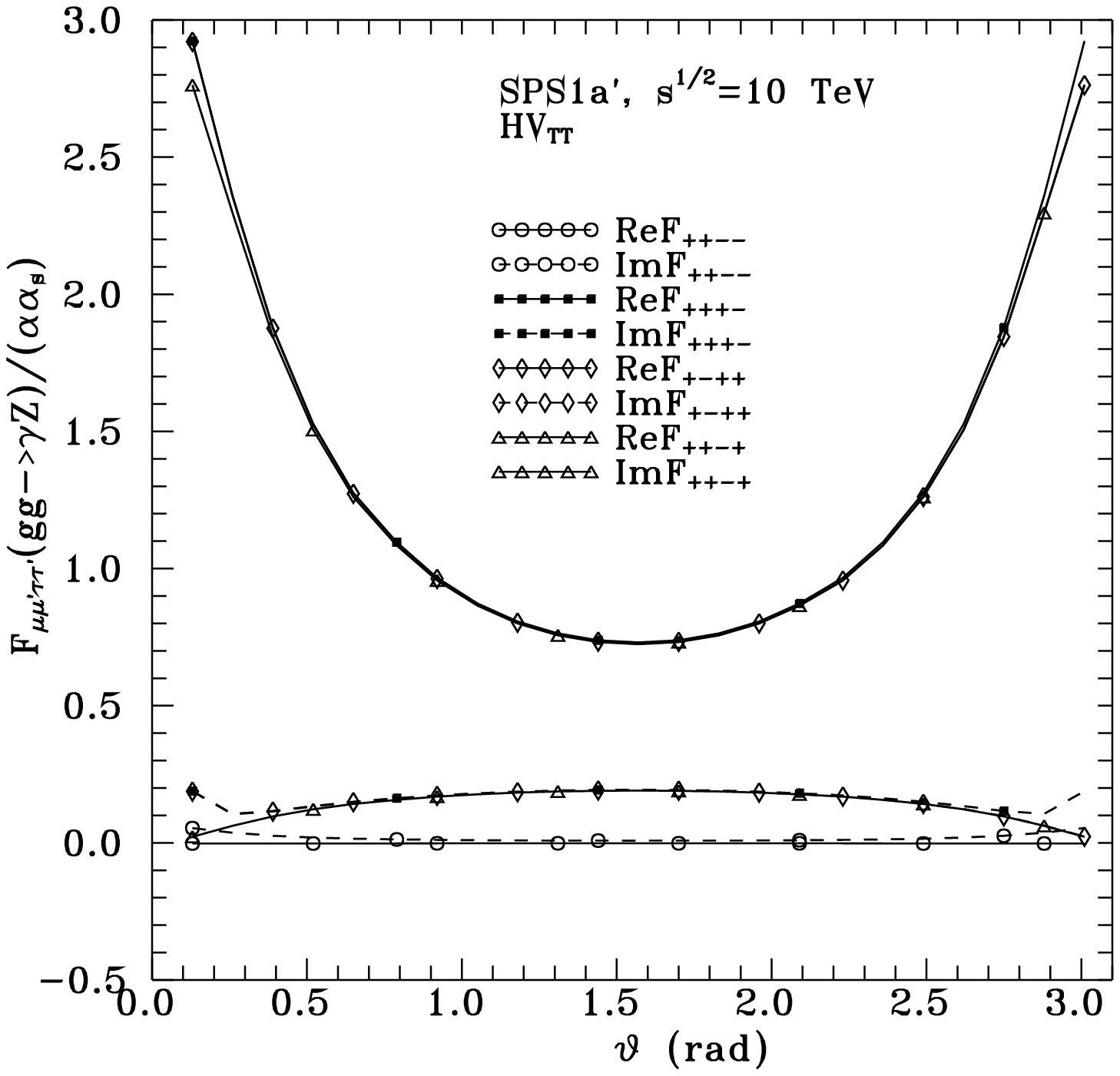, height=6.cm}\hspace{1.cm}
\epsfig{file=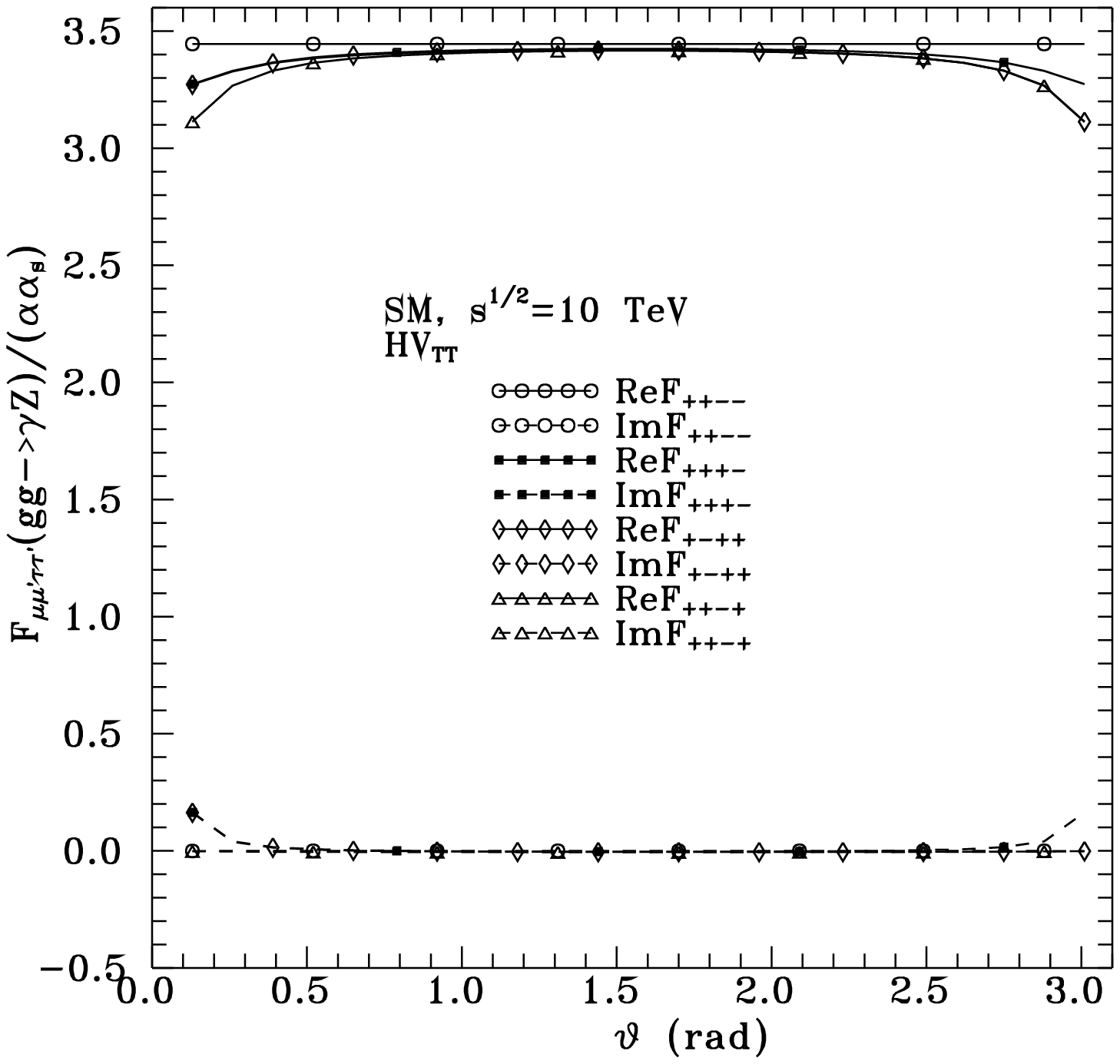,height=6.cm}
\]
\caption[1]{$HV_{TT}$-amplitudes for $gg\to \gamma Z $ in $SPS1a'$ and SM,
as in Fig.\ref{HC-ggWW-amp-fig}.  }
\label{HV-TT-gggZ-amp-fig}
\end{figure}

Using this   code we  study in detail how the 1loop EW corrected amplitudes behave,
as the energy increases. The results for these amplitudes are presented
in figures bellow, the main  goal of which is to show the differences
between SM and MSSM.   Thus,
in Figs.\ref{HC-ggWW-amp-fig},\ref{HV-TT-LL-ggWW-amp-fig}
we present  the $HC$ and $HV_{TT,LL}$
independent helicity amplitudes for $gg\to W^+W^-$, defined in  (\ref{F-ind-WW}).
The corresponding ones
for $gg\to ZZ$, are  shown in Figs.\ref{HC-ggZZ-amp-fig},
\ref{HV-TT-LL-ggZZ-amp-fig} using  (\ref{F-ind-ZZ}); while in
 Figs.\ref{HC-gggZ-amp-fig}, \ref{HV-TT-gggZ-amp-fig}, the HC and $HV_{TT}$
 independent amplitudes for $gg \to \gamma Z$ are shown using (\ref{F-ind-gammaZ}).

In all these cases, left panels  give the
$SPS1a'$ MSSM result  \cite{SPA1}, and right panels the SM one.
The upper  panels always describe the energy distributions, while
the lower panels give the angular distributions.
Note that the sensitivity of the MSSM result on the specific benchmark, is only
at intermediate energies. The high energy limit is  independent of it;
while at energies much below $M_{SUSY}$, all  MSSM amplitudes coincide with the SM ones .

No amplitudes for
$gg \to \gamma\gamma$ are shown, since their structure is very similar
to the  $gg\to \gamma Z$ results, apart  from the overall scale factors; compare
(\ref{R8F}, \ref{Casym2}).\\

As seen in Figs.\ref{HC-ggWW-amp-fig}, \ref{HC-ggZZ-amp-fig}, \ref{HC-gggZ-amp-fig},
the shapes of the various HC amplitudes in MSSM ($SPS1a'$) and SM, are rather  similar,
for all processes (\ref{processes}),   while   their asymptotic values are reached
at energies  $\lsim  4 TeV$.

The differences between MSSM and SM become striking though, for the $HV_{TT,LL}$
amplitudes in  Figs.\ref{HV-TT-LL-ggWW-amp-fig}, \ref{HV-TT-LL-ggZZ-amp-fig},
\ref{HV-TT-gggZ-amp-fig}. In  MSSM (left panels)  they   vanish quickly like
\bqa
&& F_{+++-} \simeq F_{+-++} \sim \frac{m}{\sqrt{s}} \log^n s  ~~, \nonumber \\
&& |F_{+++-}| \gsim  |F_{++00}| \gsim |F_{++--}| \sim \frac{m^2}{s}\log ^{n'}s ~~,
\label{WW-ZZ-amp-asym}
\eqa
for  $gg \to W^+W^-, ~ZZ $; while for  $gg \to \gamma Z$ they behave as
\bq
 F_{+++-} \simeq F_{++-+}\simeq F_{+-++} > |F_{++--}|  \sim \frac{m^2}{s}\log ^{n''}s~~,
 \label{gamZ-amp-asym}
\eq
where $n,n',n''$ are numbers.

In SM (right panels of Figs.\ref{HV-TT-LL-ggWW-amp-fig}, \ref{HV-TT-LL-ggZZ-amp-fig},
\ref{HV-TT-gggZ-amp-fig}) all  $HV_{TT,LL}$  amplitudes go to constant,
angle-independent limits, as required by (\ref{Fasym1}, \ref{ASHV}, \ref{HVLLasym}).
Moreover, for $gg \to W^+W^-, ~ZZ $,  the $F_{++00}$ amplitude is  the largest one,
above $\sim 2 TeV$. This is also confirmed by Table 1.\\

We next turn to the TL  or LT amplitudes, which vanish at high energies,
in both MSSM and SM. The difference between the MSSM and SM predictions for these amplitudes
is very small, for all energies. Therefore, we only show MSSM results here.

\begin{figure}[p]
\[
\hspace{-0.5cm}
\epsfig{file=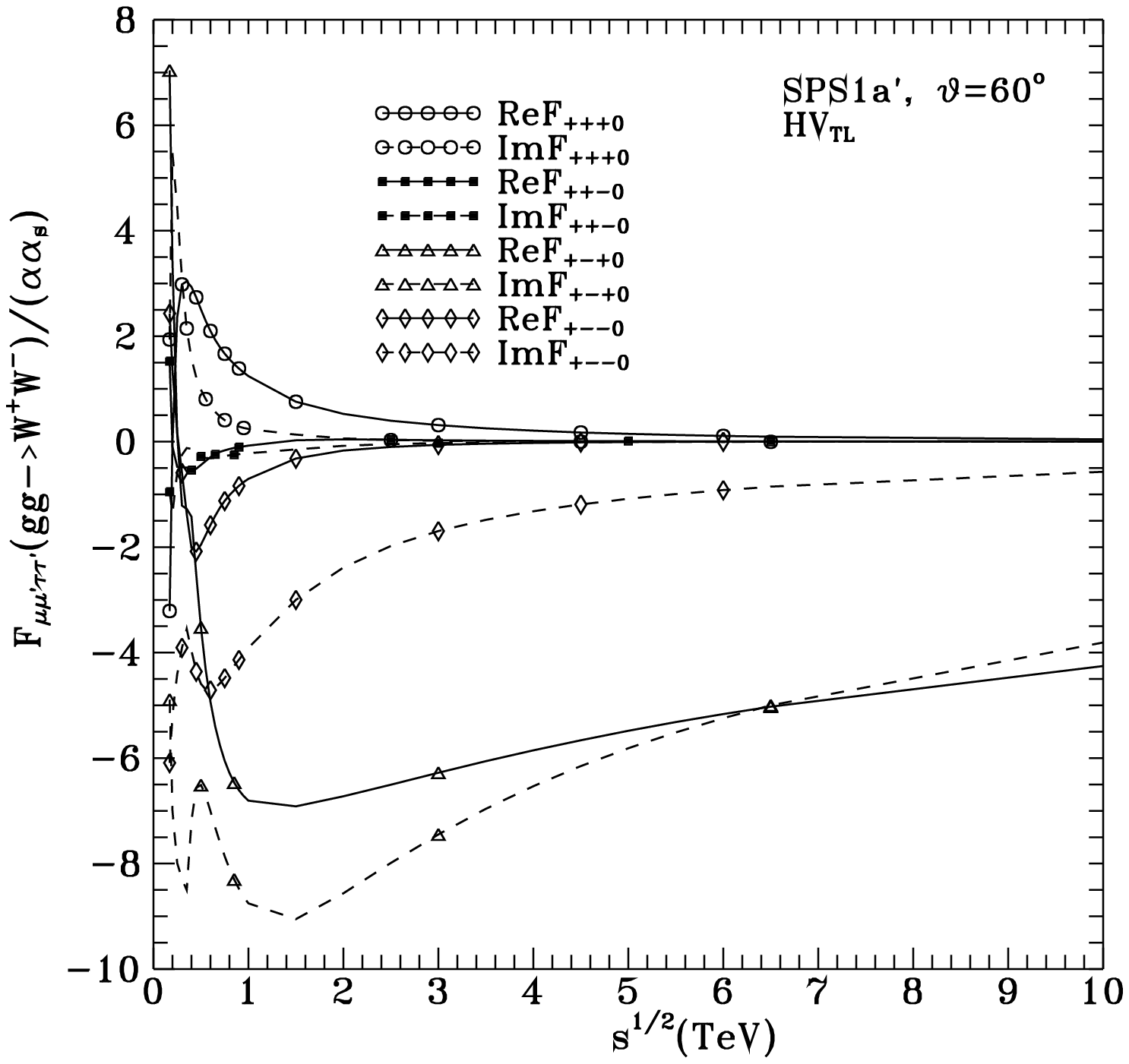, height=6.cm}\hspace{1.cm}
\epsfig{file=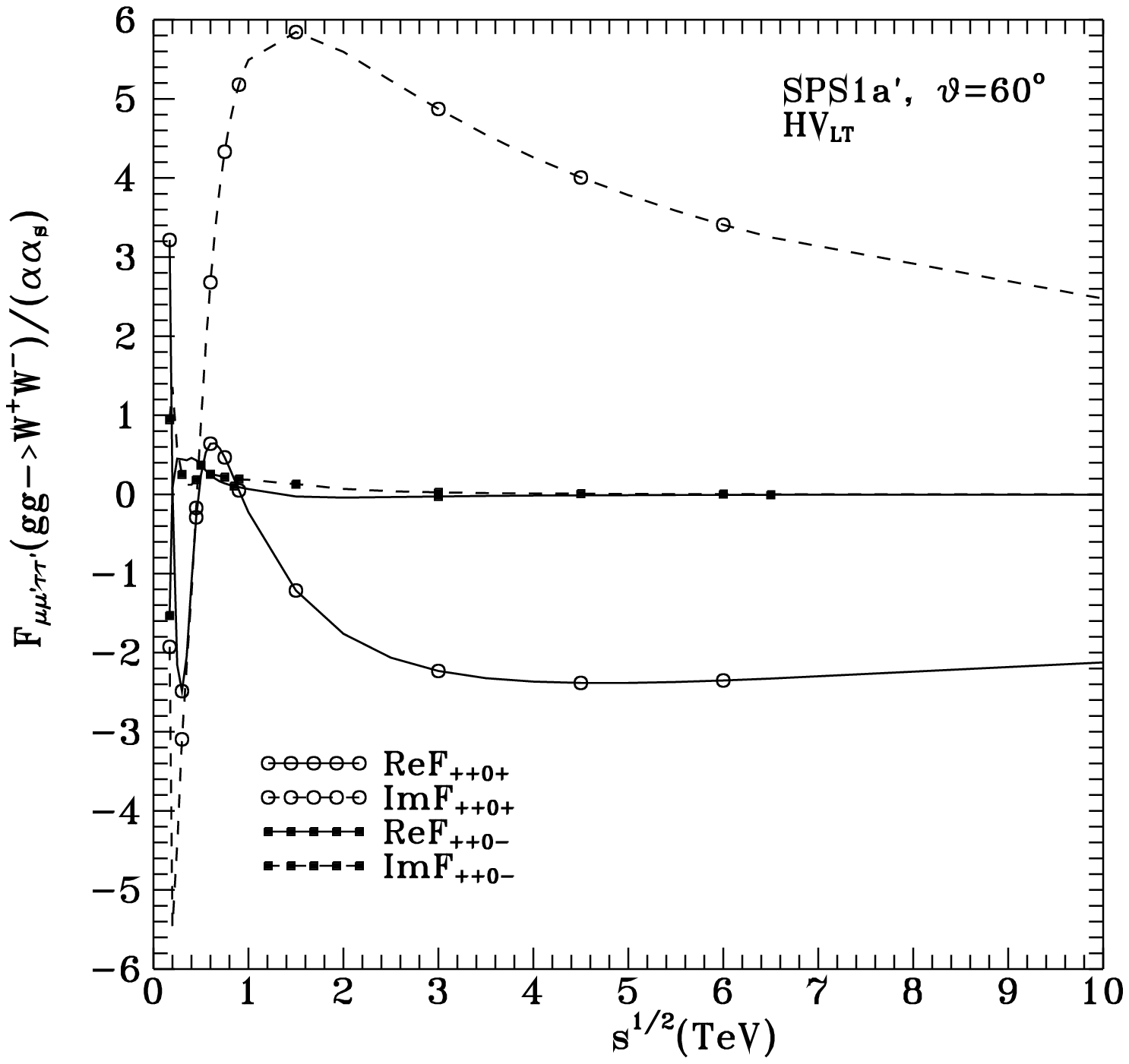,height=6.cm}
\]
\[
\hspace{-0.5cm}
\epsfig{file=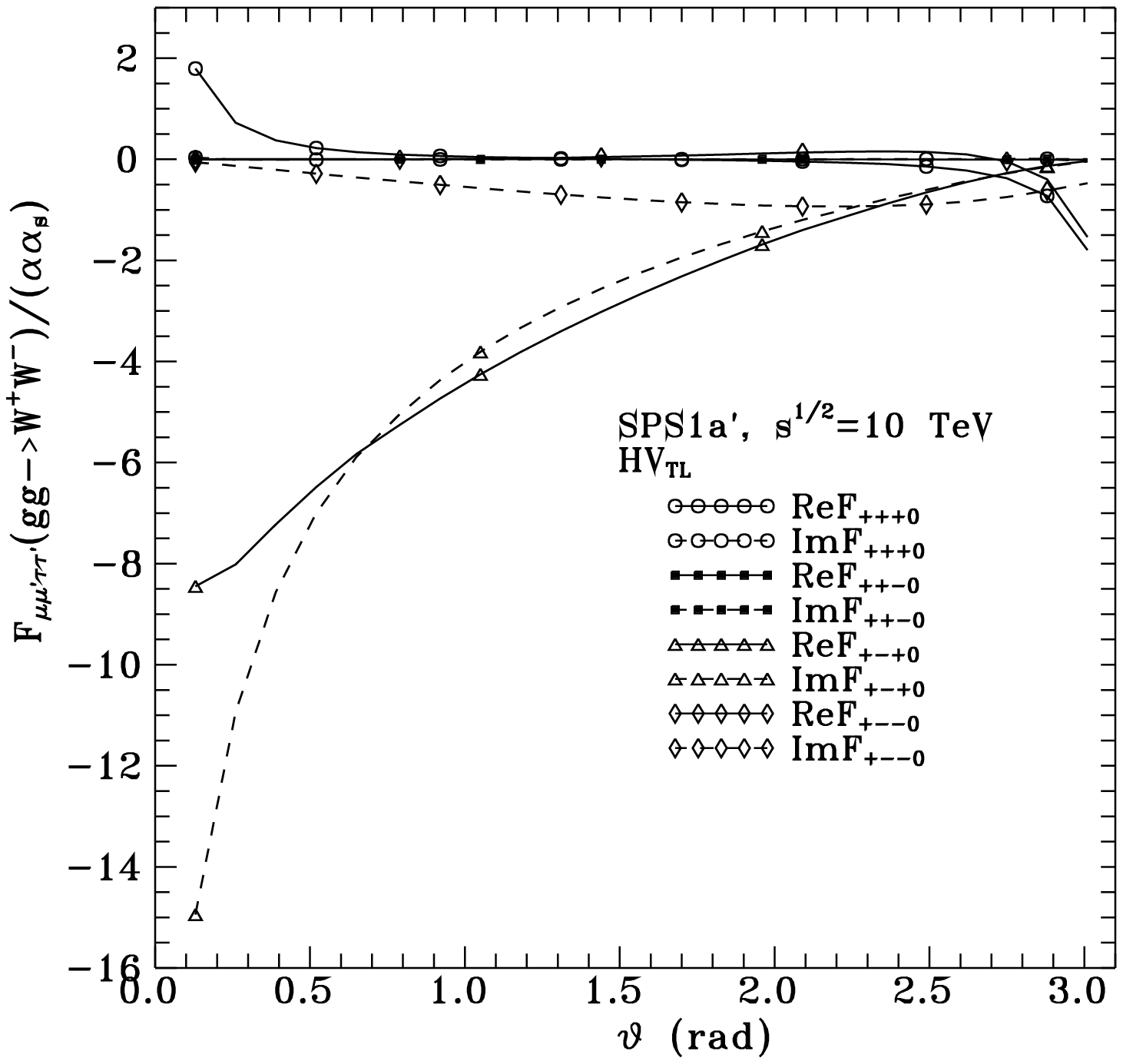, height=6.cm}\hspace{1.cm}
\epsfig{file=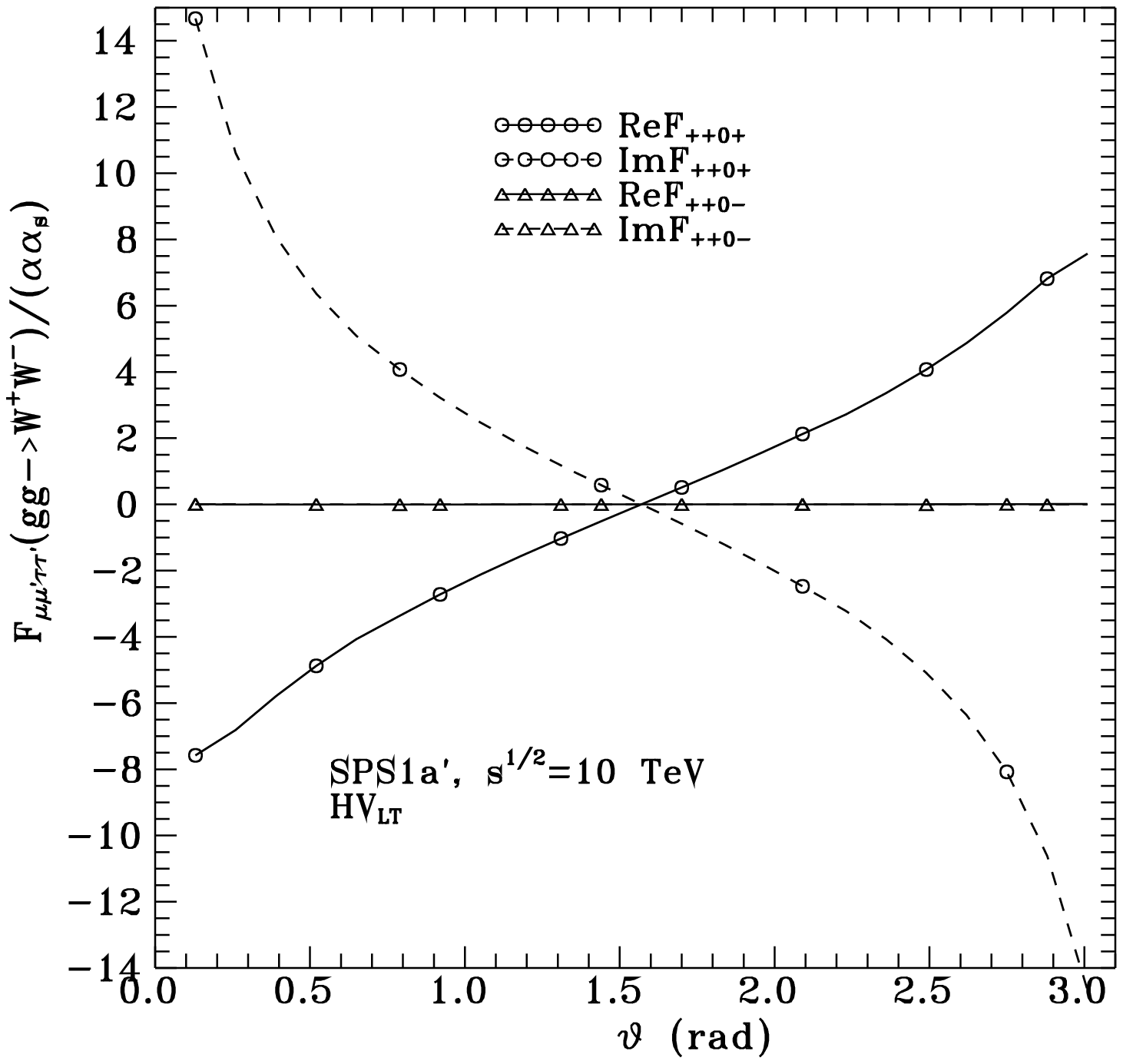,height=6.cm}
\]
\caption[1]{$HV_{TL}$ (left panels) and $HV_{LT}$ (right panels)  amplitudes
 for $gg\to W^+W^- $ in $SPS1a'$; $\tau \tau'=0$. Upper panels give the energy
 distributions at $\theta=60^o$, while the lower panels give the angular distribution
 at $\sqrt{s}=10 TeV$.  }
\label{HV-TL-ggWW-amp-fig}
\end{figure}

In  Figs.\ref{HV-TL-ggWW-amp-fig}
we  present the  TL (left panels)  and LT (right panels) amplitudes
for $gg \to W^+W^-$ for  $(SPS1a')$.
The upper panels give the energy distributions, while the lower panels give
the angular distributions at $\sqrt{s}=10TeV$. As seen there, all these amplitudes
vanish quickly, with increasing energy; while
$F^{WW}_{++0-}(\theta)\simeq - F^{WW}_{++-0}(\theta)\simeq 0 $ at almost all energies.

\begin{figure}[p]
\vspace*{-1cm}
\[
\hspace{-0.5cm}
\epsfig{file=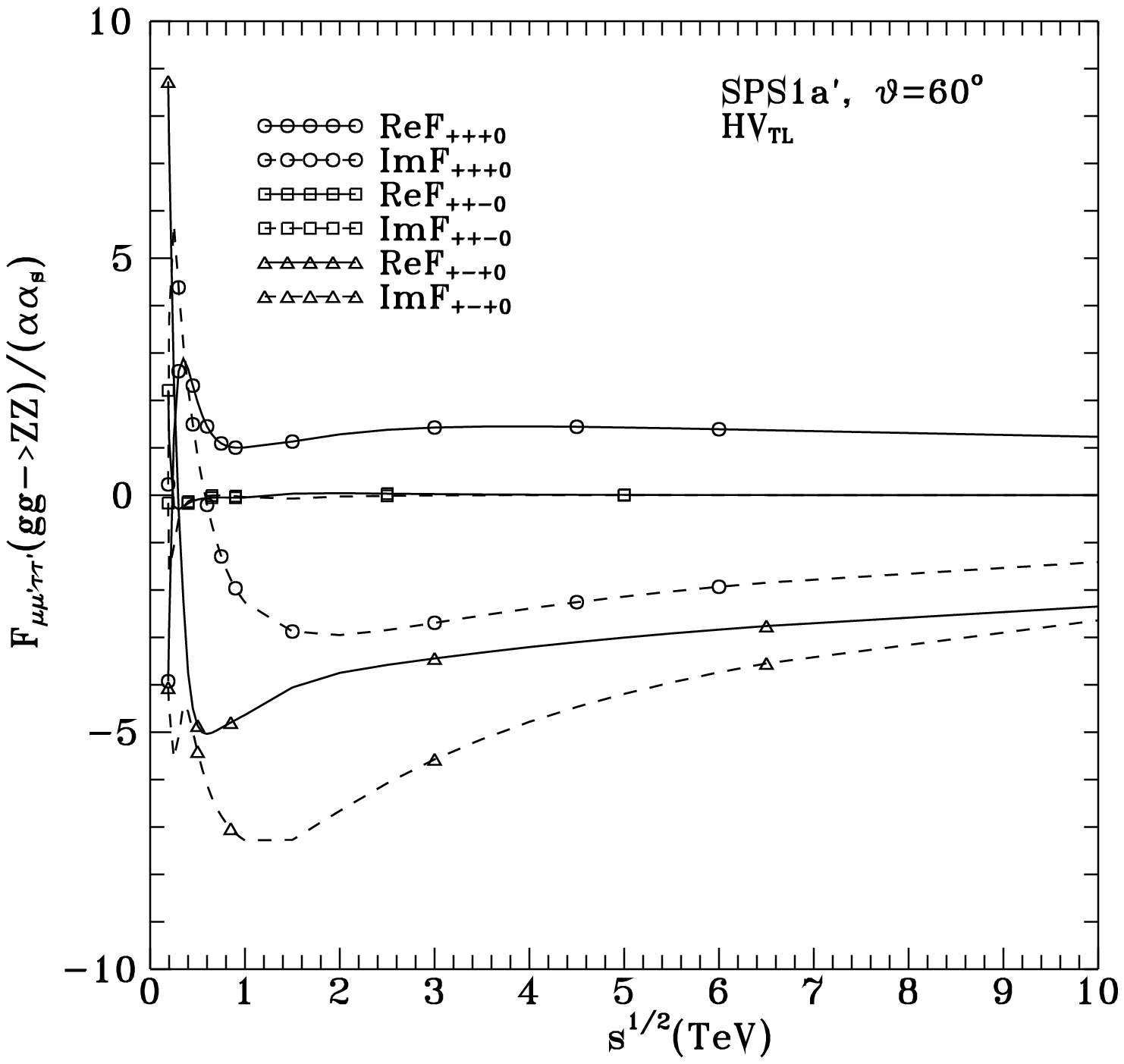, height=6.cm}\hspace{1.cm}
\epsfig{file=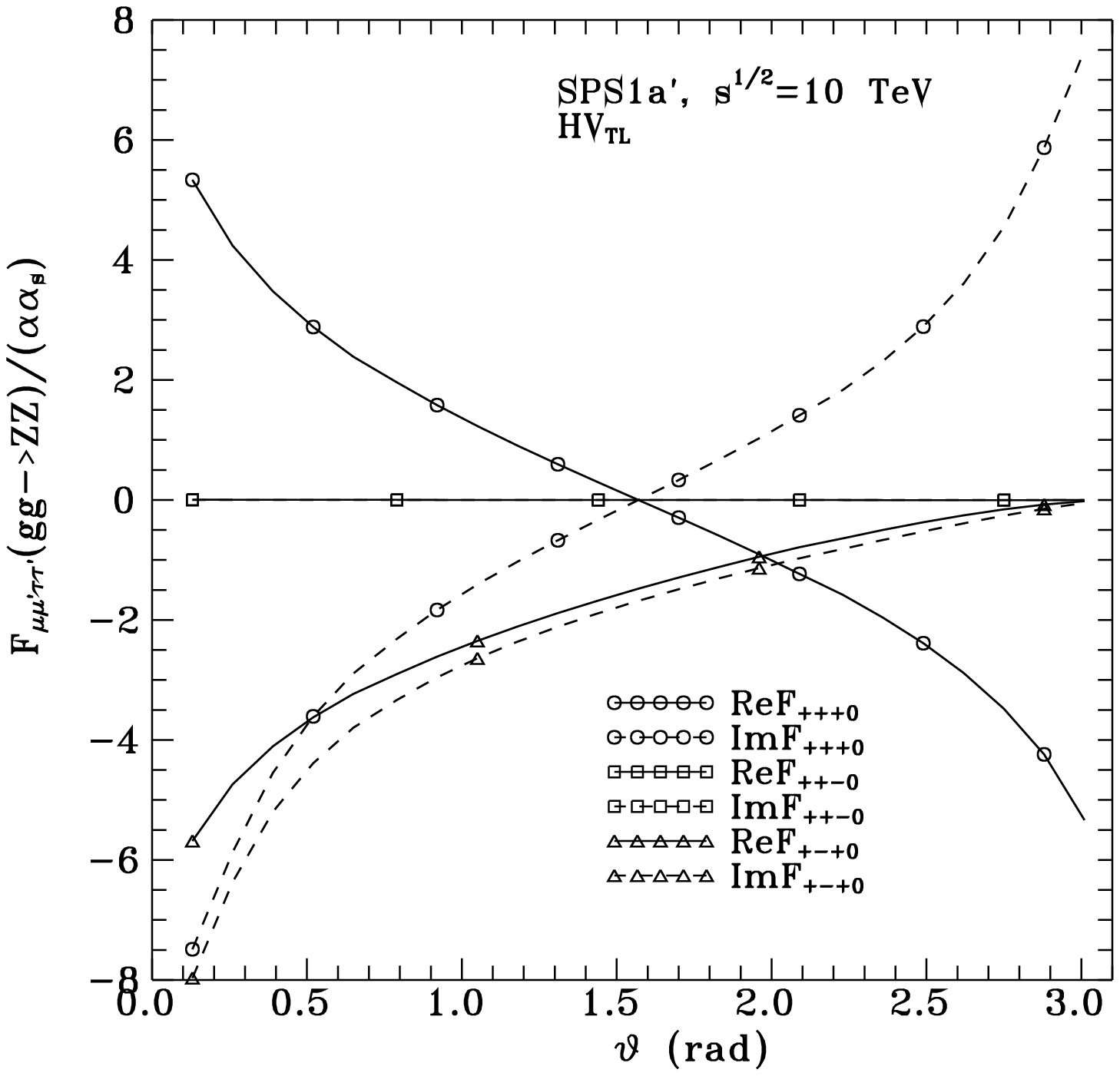, height=6.cm}
\]
\caption[1]{$HV_{TL}$  amplitudes for $gg\to ZZ $ in $SPS1a'$.
Left panel gives the energy distribution at $\theta=60^o$, while the right
panel gives the angular distribution at 10TeV.  }
\label{HV-TL-ggZZ-amp-fig}
\end{figure}

A similar behavior is observed in Figs.\ref{HV-TL-ggZZ-amp-fig} presenting
the  $HV_{TL}$ amplitudes for $gg \to ZZ$, with the left panel giving the
energy distributions, and the right panel the angular ones, always in MSSM.
Again $F^{ZZ}_{++0-}(\theta) = - F^{ZZ}_{++-0}(\theta)\simeq 0 $
is found at almost all energies. The corresponding amplitudes
for $gg \to \gamma Z$ vanish even faster, as the energy increases.

Concerning  the angular dependencies of the various  amplitudes
 appearing in Figs.\ref{HC-ggWW-amp-fig}-\ref{HV-TL-ggZZ-amp-fig},
we remark that their approach to asymptopia  for the MSSM $(SPS1a')$ benchmark,
 reaches the 10\%  level, already at an energy of about 10 TeV, for a wide angular region.
 This is most convincingly seen by comparing the angular distributions
 for the dominant HC amplitudes in Figs.\ref{HC-ggWW-amp-fig}, \ref{HC-ggZZ-amp-fig},
 \ref{HC-gggZ-amp-fig}, with the suppressed HV amplitudes in
 Figs.\ref{HV-TT-LL-ggWW-amp-fig}, \ref{HV-TT-LL-ggZZ-amp-fig},
 \ref{HV-TT-gggZ-amp-fig}, \ref{HV-TL-ggWW-amp-fig}, \ref{HV-TL-ggZZ-amp-fig}.\\

\begin{figure}[p]
\vspace*{-1cm}
\[
\hspace{-0.5cm}
\epsfig{file=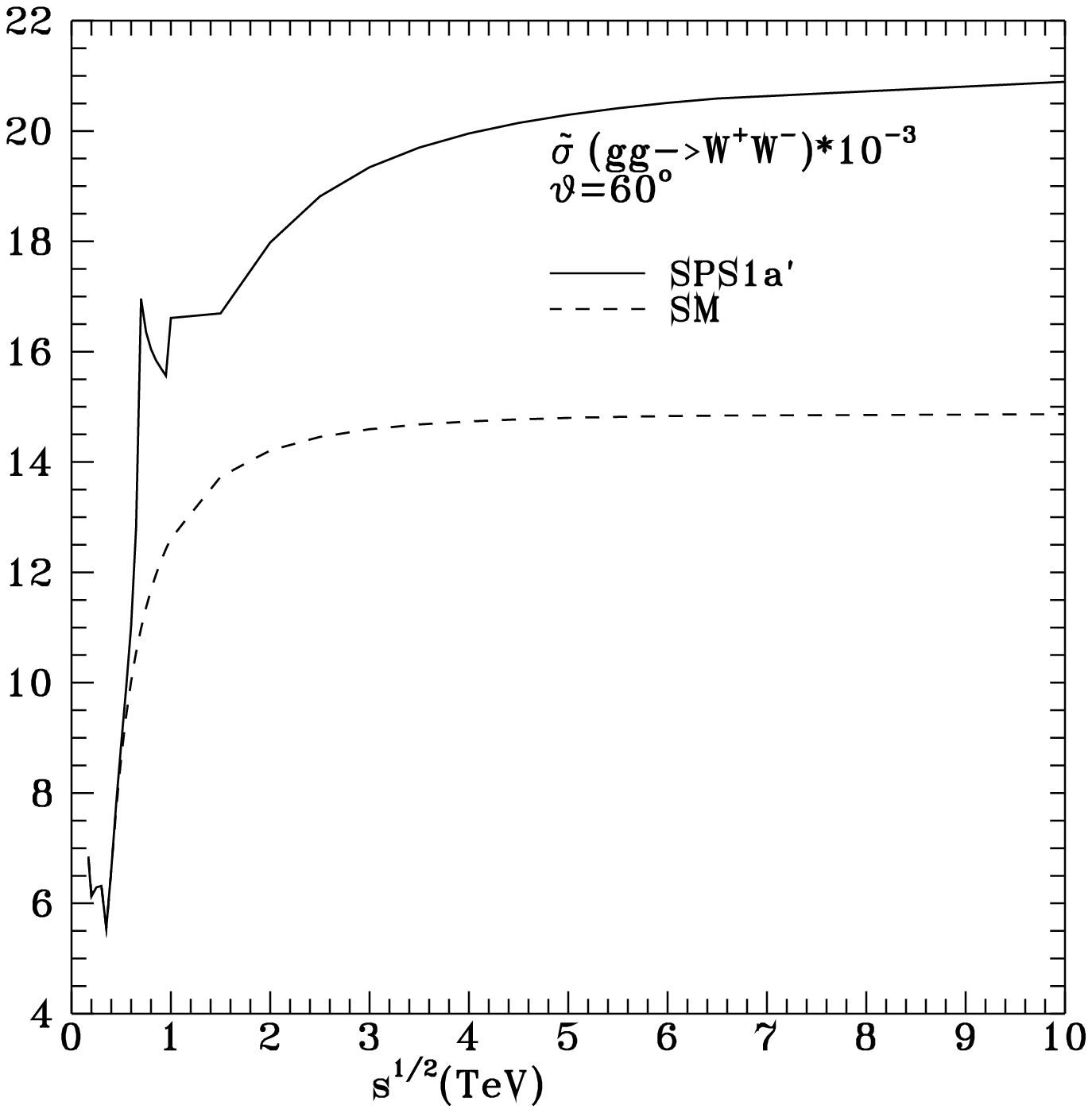, height=6.cm}\hspace{1.2cm}
\epsfig{file=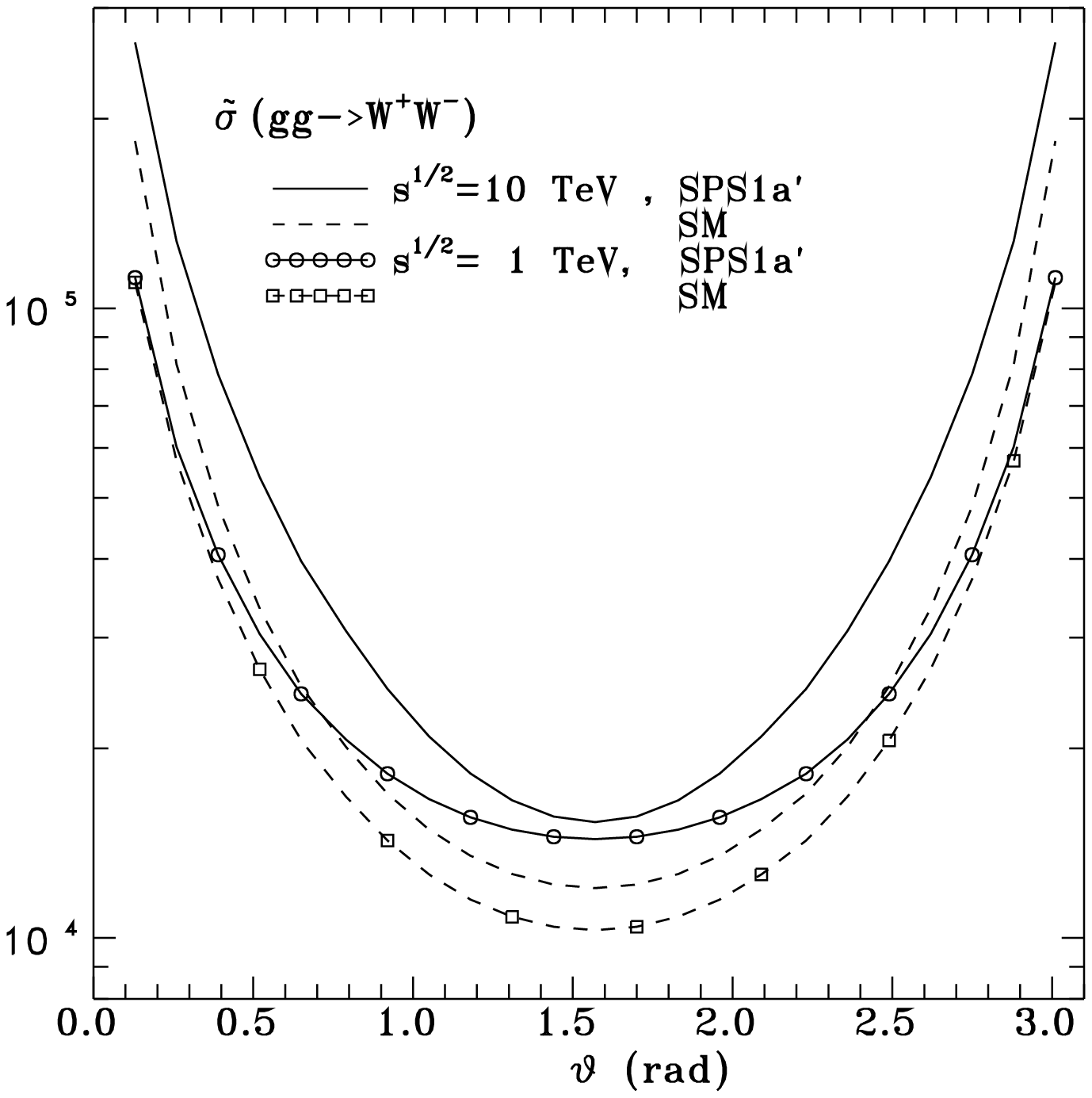,height=6.cm}
\]
\[
\hspace{-0.5cm}
\epsfig{file=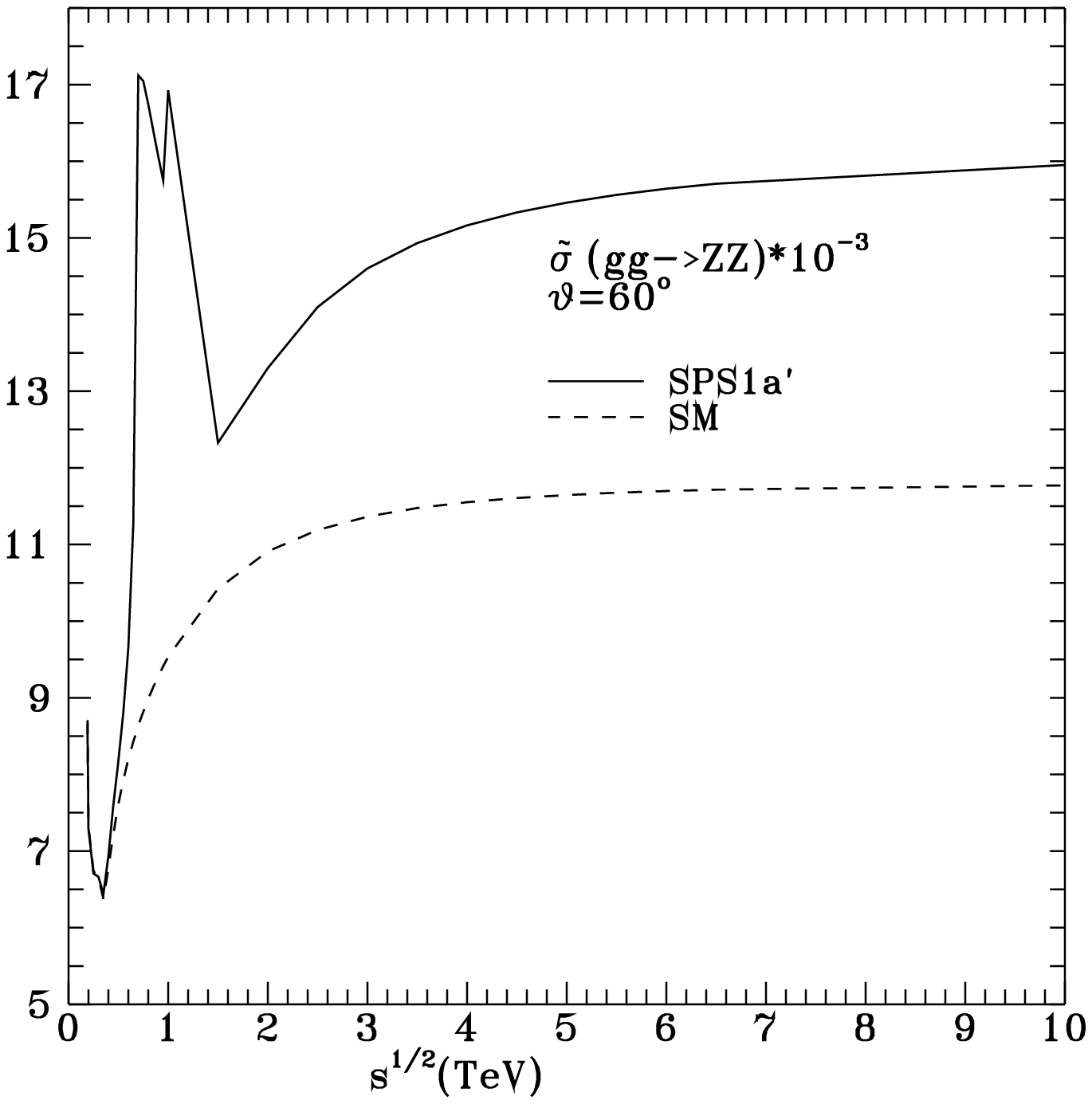, height=6.cm}\hspace{1.2cm}
\epsfig{file=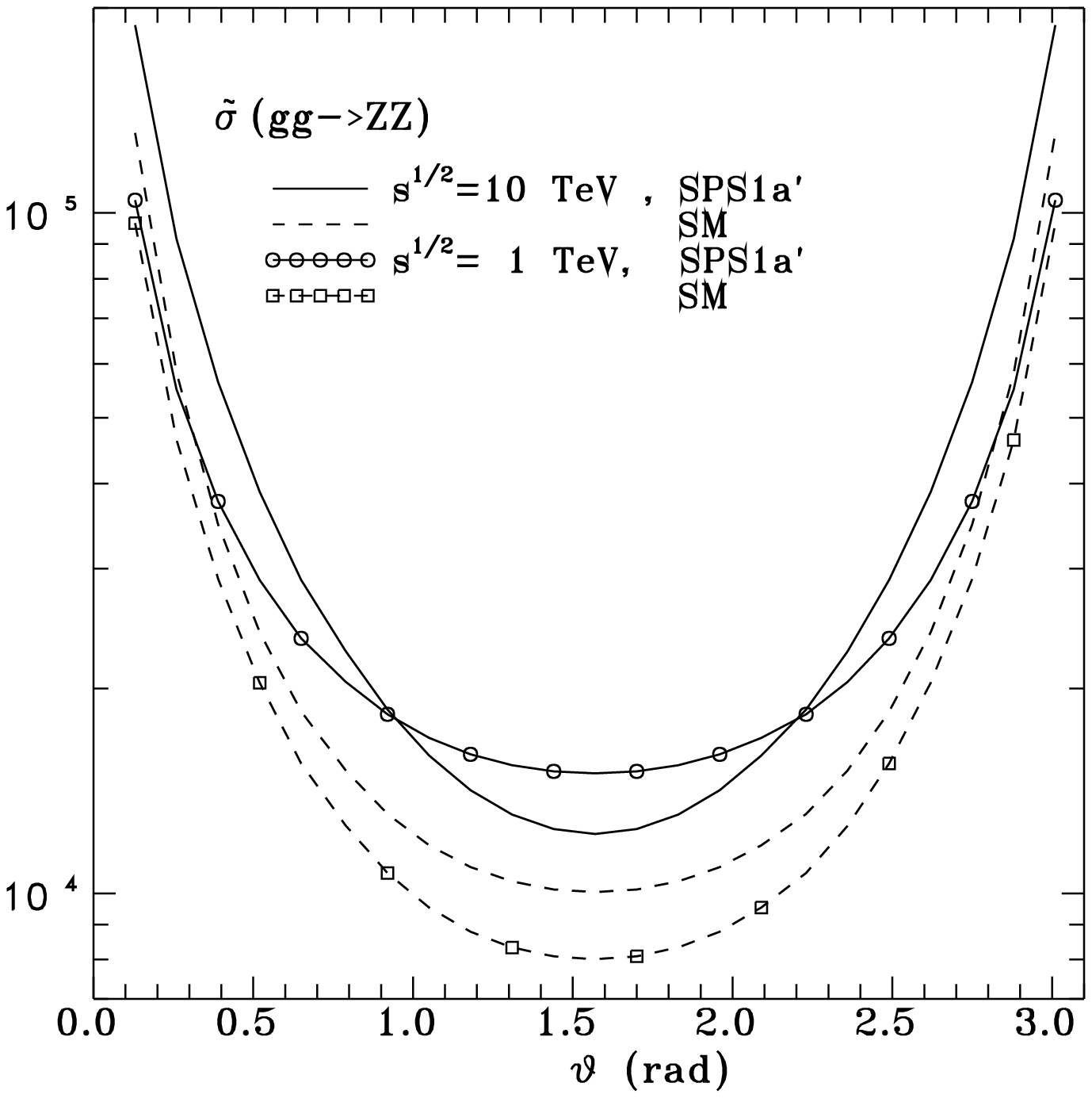,height=6.cm}
\]
\[
\hspace{-0.7cm}
\epsfig{file=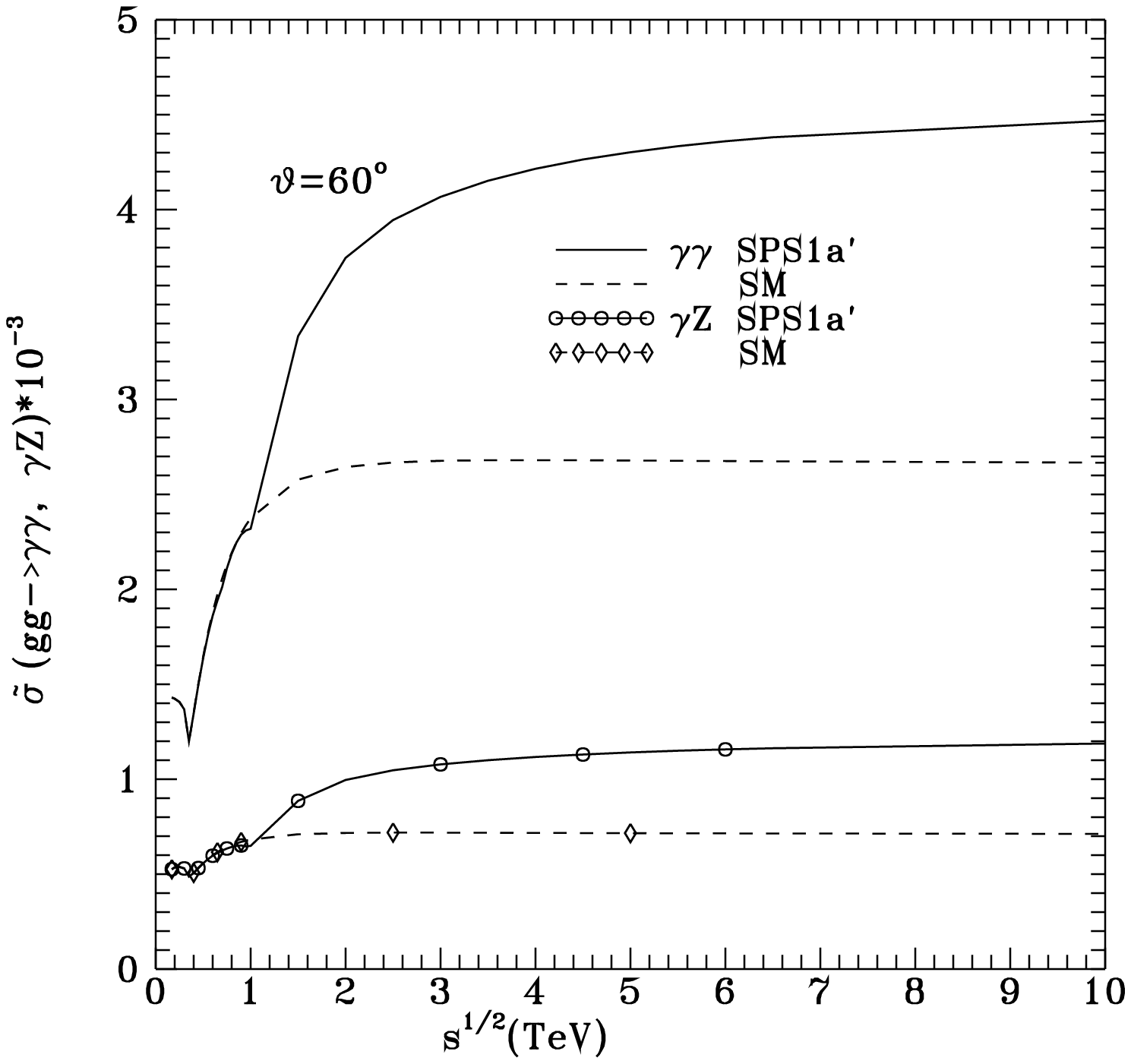, height=6.cm}\hspace{1.cm}
\epsfig{file=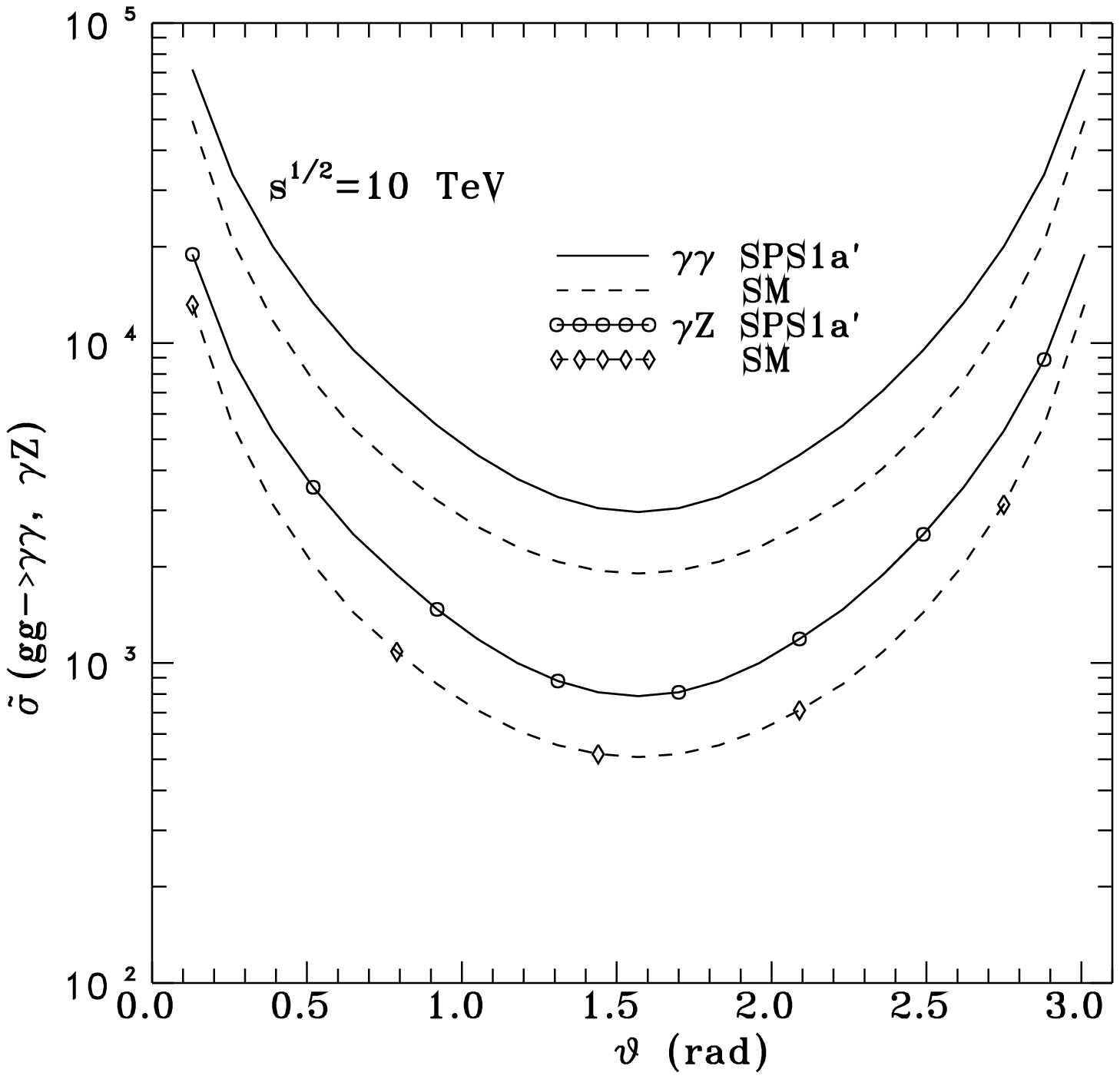,height=6.cm}
\]
\caption[1]{Dimensionless unpolarized differential cross sections
for $\tilde \sigma(gg\to W^+W^-,~ ZZ,~\gamma\gamma, ~\gamma Z)$  defined in
(\ref{sigma-tilde}),
for  $SPS1a'$ and SM. Left panels give  the energy distributions, while right panels
the angular ones. }
\label{ggVV-sig-fig}
\end{figure}

In Fig.\ref{ggVV-sig-fig} we  present the "dimensionless" unpolarized differential
cross sections defined in (\ref{sigma-tilde}) for all four processes in (\ref{processes}).
The panels of the first row refer to the $WW$-channel,
those of the middle one to the $ZZ$-channel, and finally those of the third row to the
$\gamma Z$ and $\gamma \gamma$ channels.
In each row, the left panels give the energy dependence for a fixed angle $\theta=60^o$,
in both $SPS1a'$ of MSSM \cite{SPA1} and
in SM. Correspondingly, the right panels give the angular dependencies in radians,
 at c.m. energies 1 and 10TeV  for $WW$ and $ZZ$, but only at 10TeV
 for $\gamma\gamma$ and $\gamma Z$.

At very high energies ($s \gg M_{SUSY}^2$), the cross sections  in Fig.\ref{ggVV-sig-fig},
only   depend on the gauge couplings and the top and bottom masses;
compare the asymptotic results in the Table.
 Nevertheless, the differences between the SM and MSSM predictions,
 even at very high energies,
 are  quite large and of course  independent of the MSSM  benchmark.
At $s \ll M_{SUSY}^2$, the SM and MSSM results should of course coincide.
  The only possible benchmark dependence  may appear at energies
  comparable to  $M_{SUSY}$, which of course, apart from the actual dependence on it,
 may also push the validity of   the asymptotic predictions to higher
 or lower energies \cite{SPA1, SPA2, other-bench1, other-bench2, other-bench3}.

We are presently investigating other such  asymptotic  relations involving
the unpolarized cross sections $\tilde \sigma(gg\to VV')$,
 $\tilde \sigma(gg\to \tchi^+_i\tchi^-_j)$ and $\tilde \sigma(gg\to \tchi^0_i\tchi^0_j)$
 \cite{ggchichi}. Some of  these relations are found
to be approximately correct even at LHC type energies,  for $SPS1a'$ \cite{SPA1}
and similarly low SUSY scale benchmarks;
much like it appeared  for $ug \to d W^+, ~\tilde d_L \tchi^+_i$  in  \cite{ugsdWino}.
There exist cross section relations though, like those involving the production
of  specific neutralino or chargino pairs, where much higher energies are needed for
them  to hold \cite{ggchichi}.\\

We next turn to   what could eventually be the LHC implications of the extensive
1loop EW calculation for $gg \to VV'$, we have performed.
To avoid any misunderstanding though, we emphasize that
the reason we have performed this  calculation was not in order to check
its observability at LHC; but in order to improve our understanding
of the beautiful helicity conservation property endowed to  SUSY.

Having said this, the  contribution to  $VV'$ production due to gluon fusion at LHC,
may be written as
\bqa
&& \frac{d\sigma(pp\to VV'...)}{d s d\cos\theta} =  \frac{L_{gg}}{S}\,
\frac{d\sigma(gg\to VV' ;s)}{d\cos\theta} ~~, \nonumber \\
&& L_{gg} =  \int_0^1 \frac{dx_1}{x_1}  g(x_1, Q^2) g\left (\frac{\tau}{x_1}, Q^2\right )
~~, ~~~ \tau=\frac{s}{S}~~~, \label{LHCdsig}
\eqa
where  $g(x, Q^2)$ is the gluon  distribution function of the proton
at an appropriate scale $Q$ \cite{glumi},
$s$ is the subprocess c.m. energy-squared,
$\theta$ is the c.m. scattering angle, and
  $S$ the  LHC  energy-squared. The relation of the
  $F^{VV'}_{\mu\mu'\tau\tau'}$ amplitudes
  to the cross sections appears in (\ref{sigma-tilde}).

\begin{figure}[p]
\vspace*{-1cm}
\[
\hspace{-0.5cm}
\epsfig{file=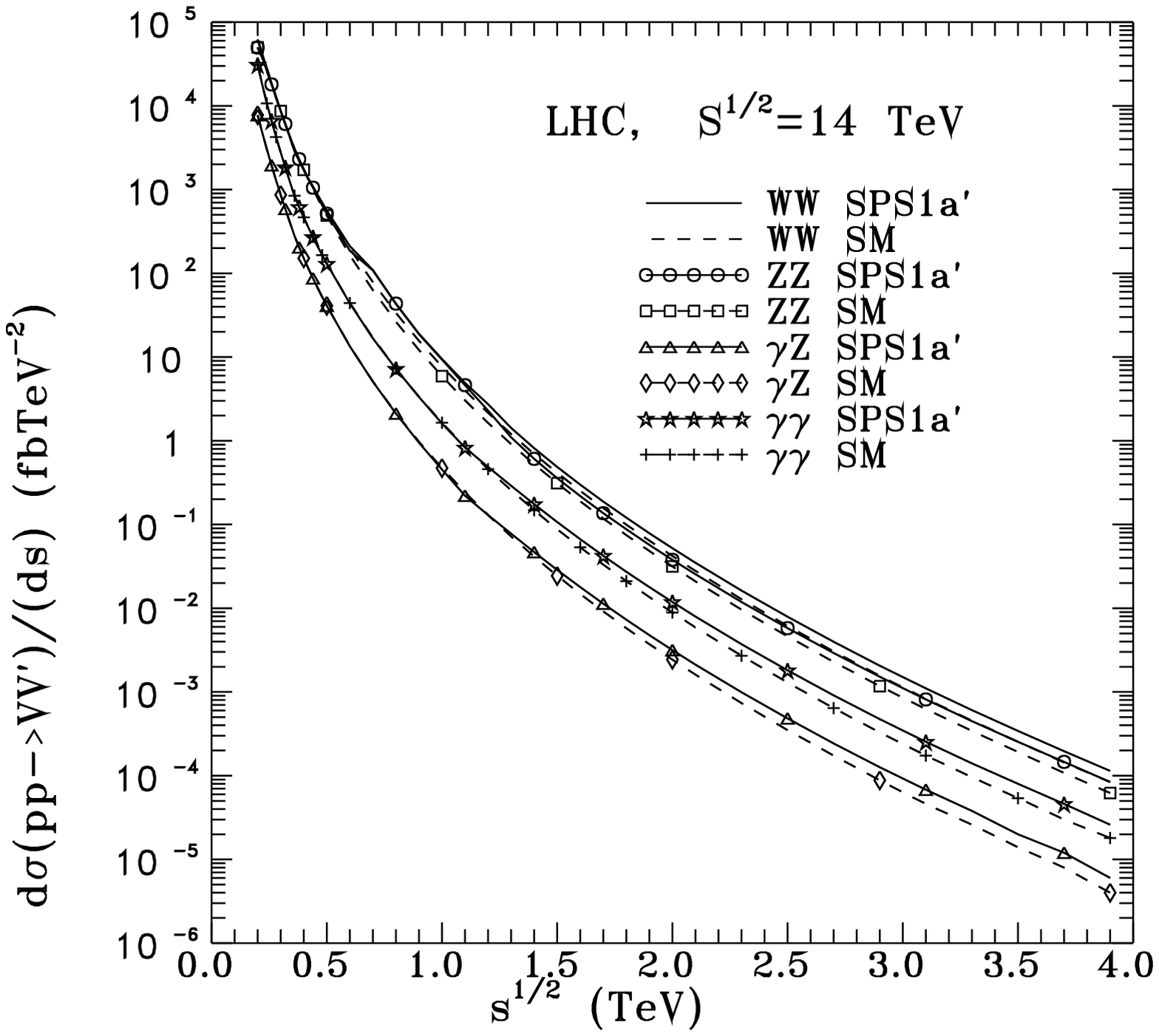, height=6.cm}\hspace{1.cm}
\epsfig{file=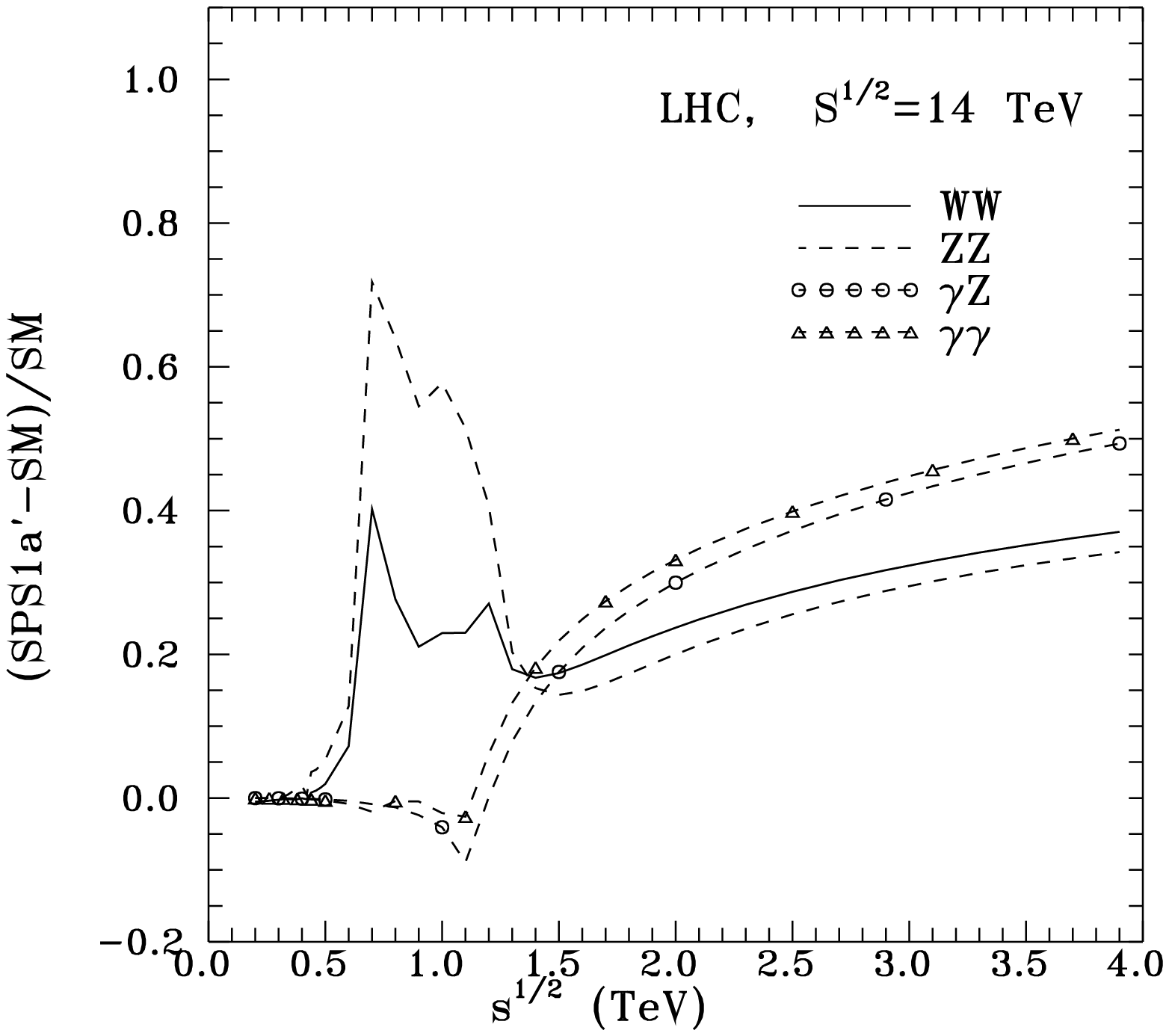, height=6.cm}
\]
\caption[1]{Left panel: The gluon fusion contribution to the LHC cross section as a function
of the   subprocess energy-squared $s$,
for  $W^+W^-$, $ZZ$, $\gamma Z$ and $\gamma \gamma$ production
in $SPS1a'$ of MSSM and in SM. Right panel: The  $SPS1a'$ result is compared to the SM one,
for all these processes.   }
\label{LHC-fig}
\end{figure}

Relation (\ref{LHCdsig}) is the basic
quantity determining  the difference between  SM
and MSSM. Integrating for example over all angles, we get
the invariant mass squared distributions $d\sigma(pp\to VV')/ds$  presented in
the left panel of Fig.\ref{LHC-fig}, for $SPS1a'$ \cite{SPA1} and SM.
In it we used $Q=\sqrt{s}$, and the MRST2006nnlo code for the gluon distribution
function\footnote{Very similar results are expected for e.g the lowest order
gluon distributions MRST2004lo of \cite{glumi} also.} \cite{glumi}.

Correspondingly, in the right panel of Fig.\ref{LHC-fig}, we show the relative
changes that MSSM   creates to the SM predictions for $d\sigma(pp\to VV')/ds$.
As seen there,  the difference between the MSSM and the SM predictions at $\sim 1 {\rm TeV}$,
reaches the $30\%$ level for  the $WW$ channel, while for the $ZZ$
channel it approaches $70\%$. For the $\gamma \gamma$
and $\gamma Z$ channels, the effect is rather small.
This is mainly  due to the  triangle graphs involving  Higgs
exchange in the $s$-channel, which are very important for the $WW$ and $ZZ$
channels (due to a large  LL contribution),
but absent in the $\gamma Z$ and $\gamma \gamma$ channels.

Integrating $d\sigma(pp\to VV')/ds$  for all invariant masses higher than
0.5, 1 or 2 TeV in SM and $SPS1a'$, we obtain the results of Table 2,
indicating again an appreciable  SUSY effect in the $WW$ and $ZZ$ channels,
provided the SUSY scale is in the  $SPS1a'$-range.

\begin{table}[h]
\begin{center}
{ Table 2: Gluon fusion contribution to the LHC $VV'$ production cross section in SM,
and the relative $SPS1a'$ MSSM effects,  for various ranges
of the subprocess energy $\sqrt{s}$.}\\
  \vspace*{0.3cm}
\begin{small}
\begin{tabular}{||c||c|c||c|c||}
\hline \hline
& \multicolumn{2}{|c||}{$\sigma (pp \to \gamma \gamma)_{gg} $} &
\multicolumn{2}{c||}{$\sigma (pp \to \gamma Z)_{gg} $} \\
 \hline
 & SM (fb) & ($SPS1a'$ -SM)/SM    & SM (fb) & ($SPS1a'$ -SM)/SM     \\ \hline
 $\sqrt{s}> 0.5$ TeV & 17.4  & 0.005  & 5.33  & 0.005  \\
  $\sqrt{s}> 1$ TeV & 0.657  & 0.055  & 0.186  & 0.008  \\
 $\sqrt{s}> 2$ TeV &  0.011 & 0.37  & 0.003  & 0.341  \\
   \hline \hline
& \multicolumn{2}{|c||}{$\sigma (pp \to ZZ)_{gg} $} &
\multicolumn{2}{c||}{$\sigma (pp \to W^+W^- )_{gg} $} \\
 \hline
 & SM (fb) & ($SPS1a'$ -SM)/SM    & SM (fb)  & ($SPS1a'$ -SM)/SM     \\ \hline
 $\sqrt{s}> 0.5$ TeV & 64.9  & 0.239  & 77  & 0.125  \\
  $\sqrt{s}> 1$ TeV & 2.35  & 0.417  & 3.09  & 0.224   \\
 $\sqrt{s}> 2$ TeV & 0.039  & 0.232  & 0.052  & 0.266  \\
   \hline \hline
\end{tabular}
 \end{small}
\end{center}
\end{table}

The (20-30)\% $SPS1a'$ effects for $gg\to ZZ, ~W^+W^-$ in Fig.\ref{LHC-fig}
and Table 2, may be observable at LHC and interesting; note that the QCD and parton-distribution
uncertainties are usually less than 10\%. For concrete predictions though, a detail
collider simulation study is needed, including the potentially
much larger  $q\bar q \to VV'$ subprocess cross section and a background study,
using appropriate  LHC cuts.
 Such a study is much beyond the scope of the present work.

\section{Summary and future developments}

In the present work, we have made
a complete 1loop computation of the electroweak contributions
to the processes $gg  \to \gamma\gamma, \gamma Z, ZZ, W^+W^-$.
Since there is no Born terms to
these, their amplitudes reflect deeply  the features of the
electroweak dynamics.

These amplitudes  are divided into two classes; the helicity conserving (HC)
ones that in MSSM respect HCns, and the helicity violating (HV) ones, that violate it.
We have calculated them in SM and MSSM, and a numerical code is released,
which gives  all amplitudes
as functions of the c.m energy and angle.
All input parameters are  assumed real and taken at the EW scale \cite{code}..

In addition, simple analytic  expressions for  all  asymptotic  amplitudes
in  SM and MSSM, have been established,   which are valid up to small corrections
$\O(m/\sqrt{s})$ or $\O(m^2/s)$.  These expressions    depend only
on the gauge couplings  and the top and bottom masses.
For the HC amplitudes in particular, they are energy-independent,
but  depend on the scattering angle, in both MSSM and SM.

In MSSM, all   HV amplitudes of course vanish asymptotically, and HCns is respected.

But in SM, only the TL and LT  amplitudes tend to zero at high energies.
All the TT and LL helicity violating  amplitudes  become asymptotically real
constants, independent of the c.m. angle, and comparable in magnitude to the
HC ones. There are twelve  independent such TT amplitudes,
and two LL ones, indicating   that HCns  is strongly violated in
SM for  $gg \to \gamma\gamma,~\gamma Z, ~ZZ, ~W^+W^-$.

In \cite{ggHHVH} we have seen such a strong violation of HCns in SM,
but it only affected amplitudes related by the equivalence theorem to the two LL
above \cite{equiv1, equiv2, equiv3}. The HCns  violation
for the transverse HV amplitudes presented here, is a new result involving
many more amplitudes.

The HCns validity in MSSM  for transverse amplitudes were previously seen in
$\gamma\gamma \to \gamma\gamma, ~\gamma Z, ZZ$ processes,
\cite{gamgamgamgam, gamgamgamZ, gamgamZZ}; but in that case, the $W$ loop dominance
allowed  one to infer  that helicity conservation is approximately
true in SM also. An inference  supported also by all processes
enjoying a non-vanishing Born contribution.

It is only in the present study of $gg \to \gamma\gamma,~\gamma Z, ~ZZ, ~W^+W^-$,
that a strong violation of HCns has been seen in SM, for amplitudes involving transverse
electroweak bosons.

As seen  from the figures, at very high energies the SM and MSSM amplitudes
and cross sections
are considerably or even spectacularly  different.
In fact, at such energies the difference between SM and MSSM
solely arises from the quantum numbers of the virtual particles participating in the
loops; quarks in SM, and quarks+squarks in MSSM, with their masses
and mixings being irrelevant. So $gg\to VV'$ at very high energy simply
 measure the  quark and squark degrees of freedom, and can only depend
 on the gauge couplings, as we have found above.

Intermediate energies comparable to the SUSY scale, is the place where the
differences between SM and MSSM predictions depend also  on the SUSY
masses and mixing parameters.
These differences originate from all possible amplitudes, TT, LL and TL or LT.
They affect both, the magnitude and the angular distribution
of the cross sections. The ggvvcode released here, may be used to study  such
effects for gluon fusion to EW gauge bosons.

When some SUSY masses will hopefully be discovered at LHC, the above  code, as well as the
corresponding one released in \cite{ggHHVH},  may be  used to
indicate how  the actual $gg\to  VV', ~HH', ~VH$ amplitudes compare to their asymptotic
helicity conserving values. In fact, it is through
 such  codes, and the corresponding ones released in \cite{ugdW, ugsdWino},  that
the non-asymptotic implications of the physics leading to the HCns  theorem
       \cite{heli1, heli2},   may be assessed.   \\

A nice way to understand the meaning of HCns  \cite{heli1, heli2}, is to relate it to the old
 Coleman-Mandula theorem \cite{CM}, which was claiming   that any attempt
to combine no trivially the Poincar\'{e} and internal symmetries, would necessarily lead
to vanishing amplitudes for all processes. The only known way to evade this theorem
is supersymmetry; but not without a  price. And the price SUSY has   payed, is to have most
of its 2-to-2 amplitudes  vanishing asymptotically\footnote{The "asymptotic"
specification is easily understood, since we must avoid the effects that softly
break the symmetry.}. Only the few helicity conserving
ones  can survive in this limit \cite{heli1, heli2}.

Up to now, in all our MSSM studies, we have assumed R-parity conservation.
It should be straightforward to extend HCns
to any  non-minimal   SUSY model in  four dimensions,  provided renormalizability is
respected. Renormalizability seems necessary for the validity of HCns.
All known non-renormalizable couplings violate it, already
at the tree level \cite{Kasimierz}.
The validity of HCns in cases where  R-parity is violated, needs to be investigated.

\vspace*{0.5cm}
\noindent
{\bf Acknowledgements}
GJG is  partially supported by the European Union
 contract MRTN-CT-2006-035505 HEPTOOLS, and the European
 Union ITN programme "UNILHC" PITN-GA-2009-237920. Discussions with G. Altarelli
 are gratefully acknowledged.


\end{document}